# Low temperature Terahertz Spectroscopy of LaFeO$_3$, PrFeO$_3$, ErFeO$_3$, and LuFeO$_3$: Quasimagnon resonances and ground multiplet transitions

Néstor E. Massa,*,1  Leire del Campo,2  Vinh Ta Phuoc,3  Paula Kayser,4,‡ and José Antonio Alonso5

1 Centro CEQUINOR, Consejo Nacional de Investigaciones Científicas y Técnicas, Universidad Nacional de La Plata, Bv. 120 1465, B1904 La Plata, Argentina.

2Centre National de la Recherche Scientifique, CEMHTI UPR3079, Université Orléans, F-45071 Orléans, France.

3Groupement de Recherche Matériaux Microélectronique Acoustique Nanotechnologies-UMR7347 CNRS, Université de Tours, INSA CVL, Parc Grandmont, F-37200, TOURS, France.
.

4Centre for Science at Extreme Conditions and School of Chemistry, University of Edinburgh, Kings Buildings, Mayfield Road, EH9 3FD Edinburgh, United Kingdom.

5Instituto de Ciencia de Materiales de Madrid, CSIC, Cantoblanco, E-28049 Madrid, Spain.

‡Present address: Departamento de Química Inorgánica, Facultad de Ciencias Químicas, Universidad Complutense de Madrid, 28040 Madrid, Spain.

Corresponding author:
*Néstor E. Massa,  e-mail: neemmassa@gmail.com

**Physics Subject Headings (PhySH)**

Research Areas

Spin wave resonances. Magnons. Magnetic anisotropy. Canted ferromagnetism., Antiferromagnetism. Ferroelectricity. Optical phonons.

Physical Systems

Ferrites. Multiferroics. Strongly correlated systems. Perovskites.

Techniques

Optically detected magnetic resonances. Crystal field theory. Terahertz spectroscopy. Fourier transform infrared spectroscopy

**DISCIPLINE**

Condensed Matter, Materials & Applied Physics

# Abstract

We report on zone center THz excitations of non-Jahn Teller $LaFeO_3$, $PrFeO_3$, $ErFeO_3$, and $LuFeO_3$ distorted perovskites under external magnetic fields up 7 T. Our measurements on low temperature-low energy absorptions of $LaFeO_3$ show quasiantiferromagnetic and quasiferromagnetic magnons at $\omega_{qAFM}$~31.4 and $\omega_{qFM}$ ~26.7 $cm^{-1}$ in the $\Gamma_4$ ($G_x$, $A_y$, $F_z$) representation with degeneracy near linearly lifted by the field. $LuFeO_3$ is characterized by zero field magnetic resonances at $\omega_{qAFM}$ ~26.3 $cm^{-1}$ and $\omega_{qFM}$ ~22.4 $cm^{-1}$ in addition to $Fe^{3+}$ Zeeman - split crystal field (CF) $6A_1$ ground transitions at ~10.4 $cm^{-1}$ triggered by subtle structural deviations induced by the Lu $4f^{14}$ smaller ionic radius at the A site. This local quasinon-centrosymmetric departure is also found in $ErFeO_3$ (Kramers $4f^{11}$ $Er^{3+}$ ($^4I_{15/2}$); $\Gamma_2$ ($F_x$, $C_y$, $G_z$) <$T_{SR}$ ~93 K), but with the ~4 $cm^{-1}$ $Fe^{3+}$ Zeeman branching strongly biased toward higher energies due to 3d-4f exchange. Magnons at $\omega_{qAFM}$ ~31.5 $cm^{-1}$ and $\omega_{qFM}$ ~21.5 $cm^{-1}$ in $ErFeO_3$ do not undergo field induced band splits but a 13-fold increase in the quasiantiferro ($\omega_{qAMF}$) /quasiferro($\omega_{qAFM}$) intensity ratio. There is a remarkable field-dependent CF matching population balance between $Fe^{3+}$ higher and $Er^{3+}$ lower Zeeman branches. The $Er^{3+}$ ($^4I_{15/2}$) multiplet, at the 49.5 $cm^{-1}$, 110.5 $cm^{-1}$, and 167.3 $cm^{-1}$, coincides with external lattice mode frequencies suggesting strong lattice driven spin-phonon interactions. Far infrared absorption ratios under mild external fields reveal magnetic dependence only for those zone center phonons involving moving magnetic ions. Overall, our results support the viability of magnetic state manipulation by phonons.

Quasiantiferro- and Quasiferro- resonances in $PrFeO_3$ turn much broader as non-Kramers $Pr^3$ introduces ligand changes at the A site leading into near degeneracy the quasiantiferromagnetic mode and the lowest $Pr^{3+}$ CF transition. They merge into a single broad mostly unresolved feature at 7 T.

We conclude that low energy excitations in $RFeO_3$ (R=rare earth) strongly depend on the lanthanide ionic size thus indivisibly tied to the mechanism associated to the origin of canted ferromagnetism. In addition, minute lattice displacements also underlie considering non-centrosymmetric the most distorted $RFeO_3$ (R=rare earth). In these perovskites the changes triggered in the lattice by the smaller rare earth and the nonlinear intrinsic oxygen ion polarizability, known to drive lattice instabilities, provide grounds for interplay of ionic and electronic interactions yielding ferroelectric spontaneous polarization.

# INTRODUCTION

The search of multifunctional materials with magnetic and electronic polarizations as in ferromagnetic ferroelectric multiferroics [1] contributed in the last decades to advances in material development aiming to devices that would potentially perform more than one task [2]. $RFeO_3$ (R=rare earth) is one family of these compounds sustaining a distorted cubic perovskite lattice. Increasing the degree of octahedral tilting across the series was early recognized to accommodate structural distortions and interplays in a unique magnetic and electronic environment created by the $Fe^{3+}$ and $R^{3+}$ ions [3-6]. These cooperative exchanges may then be incorporated to developments that would change our way of life as in the proposed magnetizable concrete for road electrification aimed to vehicle "on the road" charging [7].

The rich magnetic phase diagram of $RFeO_3$ originates in the $Fe^{3+}$ and $R^{3+}$ independent magnetic sublattices containing $Fe^{3+}$ canted spins tangled to iron's magnetic order and 4f lanthanide exchanges promoting magnetoelectric couplings. The consequences of this appear as changes in phonon transverse optical-longitudinal optical (TO-LO) macroscopic field magnetic and electric dependences [8, 9], Heisenberg magnons [10], crystal field (CF) levels, and in $Fe^{3+}$ canted magnons within the spectral region from far infrared to the THz.

$RFeO_3$ ferrites, with four molecule per unit cell (Z=4), belong to the room temperature crystalline space group Pbmn ($D_{2h}^{16}$) [11] with $Fe^{3+}$ $3d^5$ in a S=5/2 spin moment high state configuration. Below their Nèel temperatures $T_N$ [12], $Fe^{3+}$ is centrally located in a cage made of 6 antiparallel nearest neighbors facing each other in an isotropic antiferromagnetic coupling. This represents the strongest magnetic interaction in the type G plane arrangement of the $\Gamma_4$ ($G_x$, $A_y$, $F_z$) representation meaning a $G_x$ antiferromagnetic ground state along the *a*-axis, an $A_y$ weak antiferromagnetism along the *b*-axis, and a $F_z$ weak canted ferromagnetism along the *c*-axis [13, 14]. Next in strength is the $Fe^{3+}$-$R^{3+}$ exchange followed by the weaker antiferromagnetic $R^{3+}$- $R^{3+}$ interaction found at low temperatures when paramagnetic rare earth moment turn ordered in an environment of complex exchanges that may include exchange mediated out of ___ac___ plane magnetic components as for in the $\Gamma_1$ ($A_x$, $G_y$, $C_z$) representation. (Fig. S1 in the supplemental material[15])

The distinctive $F_z$ ferromagnetic component in $\Gamma_4$ ($G_x$, $A_y$, $F_z$) along the *c* axis originates in non-collinear spin canting with out of the plane deviations of the order of less than 1°. Regardless of the prevalent mechanism for the origin of canted ferromagnetism, either from Dzyalonshinkii-Moriya antisymmetric spin exchanges [16,17] or rare earth induced lattice anisotropies [18], existing $R^{3+}$ moments always have a finite exchange within $Fe^{3+}$ crystal field (CF) in the octahedral sublattice [19]. This triggers spontaneous reorientation (SR) of the iron moments when the field is just large enough to pull the net moment parallel to c axis [20], and as consequence, a non-zero *f*-$R^{3+}$–rare earth magnetic moment and *d*-$M^{3+}$ transition metal exchange reorients, on cooling, G antiferromagnetism and canted ferromagnetism from room temperature $\Gamma_4$ ($G_x$, $A_y$, $F_z$) to the lower temperature representation $\Gamma_2$ ($F_x$, $C_y$, $G_z$) [21, 22]

Low energy spin wave $Fe^{3+}$ magnetic excitations (Fig. 1) are zone center precession modes associated to canting that appear in the THz as weak absorptions when stimulated by the infrared light ac magnetic field. They are named in orthorhombic distorted perovskites, quasiantiferromagnetic ($\omega_{qAMF}$) and quasiferromagnetic ($\omega_{qFM}$) modes detectable when the oscillating frequency of the light matches the zone center frequency of spin wave cooperative motions [23]. Each resonance is coupled to an exchange mode so that their optical activity is not confined to a specified crystallographic direction and thus they should not view as normal modes [24]. One mode ($\omega_{qAFM}$) depends on anisotropies in the *ac* plane along the antiferromagnetic axis, with the spins net sum in this plane, while the other ($\omega_{qFM}$) aligns along the ferromagnetic *c* axis performing the net spin rocking motion following the motion of the sublattice spins [25,26]

Kittel [27], Nagamiya [28] and Keffer and Kittel [29] addressed these $k \approx 0$ magnon resonances with a simplifying approach in which room temperature $\Gamma_4(A_y)$ and low temperature $\Gamma_2(C_y)$ spin deviations were neglected. This results in a de facto reduction of the inequivalent number of molecules per unit cell in the magnetic structure from four magnetic sublattices of the d subsystem $M_i$ (i = 1,…, 4) and four rare-earth sublattices $m_j$ (j = 1, …, 4) to two with individual $Fe^{3+}$ magnetic moments $M_j$ (j= 1-4) bound by the equivalencies $M_1=M_3$ and $M_2=M_4$ in the so-called two-sublattice approximation. Accordingly, each magnetization from the two magnetic sublattices along narrow optical orbitals is named $M_1$ and $M_2$ and has its equation of motion linked by an exchange interaction that describes unequal precessions triggering the two resonant energy modes function of exchange and anisotropy fields in the orthorhombic lattice [29].

These collective spin excitations have been widely studied from the early days by many material research groups working with non-collinear antiferromagnets [24, 25] Among them, White et al [26] and Koshizuka et al [30, 31] reported temperature dependent Raman scattering of $YFeO_3$, $SmFeO_3$, $DyFeO_3$, $HoFeO_3$, and $ErFeO_3$ and Koslov et al pioneered THz techniques measuring $RFeO_3$ (R=Y, Tm, Dy, Gd, Ho, Er, Tb) [32]. Resonant optical pumping of f−f electronic transitions of $Dy^{3+}$ in antiferromagnetic $DyFeO_3$ was reported to strongly affect the induced magnetization dynamics that is intrinsically competing with the $Fe^{3+}$ off-resonant excitation of the subsystem spin waves. [33]

Constable et al [34] and Jiang et al [35] studied the temperature-dependent spin waves and neutron scattering and spin reorientation in $NdFeO_3$ respectively; Fu et al [36] measured spin resonances in $SmFeO_3$, and Zhang et al did it in $TmFeO_3$ using terahertz time domain spectroscopy [37] $YFeO_3$ zone center two spin wave modes have been studied up to17 T by Amelin et al [38] and Li et al [39] proposed a studied on cooperative exchange coupling of a spin ensemble in Y doped $ErFeO_3$. More recently, the spin dynamics of $TmFeO_3$ [40] and $TbFeO_3$ [41] have been addressed by neutron spectroscopies

Changing the magnetic anisotropy for $Fe^{3+}$ spins by resonant terahertz pumping of electronic orbital transitions in$TmFeO_3$ was found to trigger large amplitude coherent oscillations. [42] It has been further demonstrated that in $TmFeO_3$ coherent pulse steering of spins may be achieved by antenna THz picosecond electric fields coupling spins switching selected states by external magnetic fields,[43]

The low energy spin precession modes share the THz spectral region with transition metal CF ground transitions linked to local distortions and sublattice tilted angles that in turn also depend on less prone to modify chemical bond rare earths.

Early diffraction measurements on $RFeO_3$ (R= rare earth) ferrites [11, 44, 45] showed that the A polyhedral sites have an anomalous decrease in the number of rare earth-oxygen nearest neighbors that, deviating from the higher temperature cubic phase $AO_{12}$ cage, changes from 10 in $LaFeO_3$ to 6 in $LuFeO_3$ (Fig. 2). The reduction of the Fe-O-Fe "θ" bond angle from 157° in $LaFeO_3$ to 142° in $LuFeO_3$ implies subtle changes in orbital and spin ordering [46-50] that is mirrored in the Néel temperature depending linearly on the Fe-O-Fe superexchange angle and thus on the rare earth cation size [51].

The proven lattice ductility of distorted perovskites has profound implications for enabling B site crystal field symmetry forbidden $Fe^{3+}$ transitions of the $^6A_1$ multiplet. As we detail in the following sections, $Fe^{3+}$ ground state Zeeman split transitions is a distinctive feature of the $LuFeO_3$ THz spectra that weighs additionally to the sole spin modes picture found in $LaFeO_3$ that, having a lattice closer to cubic, prevents $Fe^{3+}$ CF transitions becoming optically active. Although both compounds nominally share the P$_{bnm}$ ($D^{2h}_{16}$) space group, the $Fe^{3+}$ site point group in $LuFeO_3$ is altered by the ligand distortions due to $Lu^{3+}$ smaller ionic radius in a cage with only net six Lu-O nearest neighbors creating non-centrosymmetric subtle deformations in $Fe^{3+}$-O ligands.

These $Fe^{3+}$ CF Zeeman branching in $LuFeO_3$ has as counterpart in the $ErFeO_3$ $Er^{3+}$ ($^4I_{15/2}$) multiplet split strongly biased due to iron - rare earth exchanges. $Er^{3+}$ ($^4I_{15/2}$) CF also overlaps with specific lattice phonons suggesting conspicuous spin-phonon interactions appearing as an induced local continuum through the Er-O lattice vibrational range in the 1T/0T far infrared ratios. Weak applied fields are also enough to hint variants in partially magnetic-driven reststrahlen associated to torsional LO modes macroscopic fields involving moving $Fe^{3+}$ ions.

While spin resonances under applied magnetic fields vary from characteristic enhancements to near CF degeneracies, we conclude that the overall complex setting, also undergoing some degree of lattice electric and magnetic coupling in a scenario prone to localized polaron formation, is a likely common to all ferrite oxides.

## SAMPLE PREPARATION AND STRUCTURAL CHARACTERIZATION

.
$RFeO_3$ (R= La, Pr, Er, Lu) polycrystalline samples were prepared by standard ceramic synthesis procedures. Stoichiometric amounts of analytical grade $Fe_2O_3$ and $R_2O_3$ powder oxides were thoroughly ground and heated in air at 1000ºC for 12 h and 1300ºC for 12 h in alumina crucibles. Then, pellets of ~1cm diameter, less than 2mm thick, were prepared by uniaxial pressing the raw powders and sintering the disks at 1300ºC for 2 h. The purity of the samples for all four $RFeO_3$ (R= La, Pr, Er, Lu) was checked by X-ray powder diffraction (XRD) collected at room temperature with Cu-Kα radiation. Shown in Fig. S2 in the Supplemental Material [15], all data were analyzed using the Rietveld method with refinements carried out with the program FULLPROF.

# EXPERIMENTAL DETAILS

Low temperature-low frequency absorption measurements in the spectral range from 3 cm$^{-1}$ to 50 cm$^{-1}$ with 0.5 cm$^{-1}$ resolution have been performed in the THz beamline of the BESSY II storage ring at the Helmholtz-Zentrum Berlin (HZB) in the low-alpha multi bunch hybrid mode.[56]

In the synchrotron low-alpha mode electrons are compressed within shorter bunches of only ~2 ps duration allowing far-infrared wave trains up to mW average power to overlap coherently in the THz range below 50 cm$^{-1}$. [57, 58]

For measurements under magnetic fields, we used a superconducting magnet (Oxford Spectromag 4000, here up 7.5 T) interfaced with a Bruker IFS125 HR interferometer (Fig. S3). Temperatures were measured with a calibrated Cernox Sensor from LakeShore Cryotronics mounted to the copper block that holds the sample in the magnet Variable Temperature Insert (VTI). Liquid helium cooled Si bolometers (4.2 K and 1.6 K from Infrared Labs) were used as detectors [15]. All measurements have been done in the Voigt configuration. We have also used ErFeO$_3$ powder embedded polyethylene pellets in the beamline facilities to identify Er$^{3+}$ multiplet and phonon field dependences in the far infrared by calculating the ratio of ErFeO$_3$ spectra under 1Tesla against that zero-field cooled at 5 K.

Far infrared near normal reflectivity spectra were taken on heating from 4 K to 300 K at 1 cm$^{-1}$ resolution with a Bruker 113V and a Bruker 66 interferometers with conventional near normal incidence geometry. Samples were mounted on the cold finger of a He- closed cycle refrigerator and a home-made He cryostat adapted to the near normal reflectivity attachment vacuum chamber of the interferometer. He cooled bolometer and a deuterated triglycine sulfate pyroelectric bolometer (DTGS) were employed to completely cover the spectral range of interest. A plain gold mirror and in-situ evaporated gold film were used for 100% reference reflectivity between 4 K and 300 K. We analyzed the reflectivity spectra using the standard procedures for multioscillator dielectric simulation. [59-61]

.

# RESULTS AND DISCUSSION

## a) LaFeO$_3$

Sharing the space group with the rest of RFeO$_3$ (R=lanthanide and Y) ferrites, four molecules per unit cell LaFeO$_3$ is the simplest compound that while closest to the cubic structure [50] still adopts the P$_{bnm}$-D$_{2h}^{16}$ orthorhombic space group [62-64].

LaFeO$_3$ G-type has spins coupled strongly antiferromagnetic below T$_N$ ~ 740 K [66] in high spin configuration Fe$^{3+}$ (half-filled 3D$^5$) S=5/2. Inelastic neutron scattering spin wave intensities were found well represented by the Heisenberg exchange interaction between nearest neighbor antiferromagnetic planes prompting a van Hove singularity at ~ 564 cm$^{-1}$ in the density of states spawning from 323 cm$^{-1}$ to 630 cm$^{-1}$ [66].

At lower frequencies, as pointed in the introduction, work on long wavelength spin waves has been mostly done under the umbrella of the so-called two sublattice approximation [30] that considers spins 1-3 and 2-4 equivalent with each sublattice magnetization, M$_1$ and M$_2$, having its own equation of motion linked to the exchange interaction describing unequal precessions. The resonant energies then are given

$$\hbar|\omega_k^\pm| = g\mu_B [H_A \cdot (2 \cdot H_E + H_A)^2]^{1/2} + g\mu_B H_0 \qquad (1)$$

where H$_A$ is an effective anisotropy field, H$_E$=2|J|zS/g$\mu_B$ (H$_{Ei}$ = $\lambda$M$_i$, $i$=1,2) with the exchange interaction J only for nearest neighbor spins, S is the spin moment of the ith and jth nearest-neighbor metal transition ions and *g* is the gyromagnetic ratio as the spectroscopic splitting factor [67]. A more quantitative treatment describing the THz active quasiferromagnetic and quasintiferromagnetic zone center spin wave resonances needs a full Hamiltonian incorporating anisotropies. The conventional spin Hamiltonian for orthorhombic ferrites [24, 25, 27, 35] is then written taking into account a single isotropic exchange constant coupling nearest neighbor transition metal spins, a single antisymmetric (canting) exchange constant, and two anisotropy constants as in

$$H_{spin} = 2J \sum_{i,j} S_i \cdot S_j + \sum_{i,j} D \cdot (S_i \times S_j) + \sum_{i,j} (K_{eff} \cdot S_i)^2 \quad (2)$$

We thus expect a relative weak resonant long wavelength pair with near equal intensity, and comparable temperature and magnetic field dependences at near the Brioullin zone center. These, Fig. 1, are given by

$$\omega_{qFM} = \{24 \cdot J \cdot S \cdot [2 \cdot (K_a - K_c) \cdot S]\}^{1/2} \quad (3)$$

$$\omega_{qAFM} = \{24 \cdot J \cdot S \cdot [6 \cdot D \cdot S \cdot \tan\beta + 2 \cdot K_a \cdot S]\}^{1/2} \quad (4)$$

with the quasiantiferromagnetic ($\omega_{qAFM}$) resonance at higher energies than the corresponding quasiferromagnetic ($\omega_{qFM}$), being $J$ the exchange integral ($J$ is positive for ferromagnets and negative for antiferromagnets) the isotropic Heisenberg constant and D the antisymmetric (DM) constant, $S$ is the spin moment of the i$^{th}$ and j$^{th}$ nearest-neighbor metal transition ions, and $K_a$ and $K_c$ are anisotropies along the **a** and **c** axes. $\beta$ is the momentum canting angle off the **ab** plane [68].

Fig. 3 (a) shows the LaFeO$_3$ quasiferromagnetic ($\omega_{qFM}$) and quasiantiferromagnetic ($\omega_{qAFM}$) zone center active resonances that, normalized by the absorption spectrum at 110 K, undergo a monotonous cooling down to 2.4 K. The lower and upper energy mode have frequency peaking at $\omega_{qFM}$=14.7 cm$^{-1}$ and $\omega_{qAFM}$=20.1 cm$^{-1}$ respectively. They are near temperature independent corroborating that by having La ($4f^0$) closed shell, avoiding extra potentially disturbing rare earth exchange, LaFeO$_3$ holds below T$_N$ G-type magnetism in the $\Gamma_4$ (G$_x$, A$_y$, F$_z$) [67,69-71] representation with canting angle α(deg.) ~0.52 [72] These resonances are found even in absence of an external field B$_0$, with energies dependences fixed by exchange and anisotropy fields. We found that our peak positions are also in agreement with lower resolution inelastic neutron scattering measurements reporting broad features at those frequencies [73].

Applying an external field B$_0$, means that both modes will no longer remain degenerate since if one resonant mode is close parallel to the applied effective field the second has the opposite sign so that the resulting picture is a diverging split [30]. We find this behavior fully reproduced in our measurements of LaFeO$_3$ quasiferromagnetic- and quasiantiferromagnetic-like modes as increasing applied field breaks their zone center degeneracy. This also reassures that our ZFC excitations have magnetic origin (Fig. S4 in the Supplemental Material [15]).

We compute the induced modifications in zone center absorptions by the applied field $B_0$ by normalizing the measured spectra using the zero-field cooled (ZFC) run (Fig. 4). These ratios, ($B_{0J}/0.0T$), yield an absolute account of the bulk field dependent spectral changes. The $\omega_{qAFM}$ magnon initially peaks in intensity and remains unresolved up to about 3 T, meaning that this narrower band is result of two superposing excitations that remain unresolved at our working 0.5 cm$^{-1}$ resolution. At higher fields it divides into two components. On the other hand, the quasiferromagnetic-like mode $\omega_{qFM}$ shows only a gentle gradual splitting and broadening above 3 T. At 1T, Fig. S4 in the Supplemental Material [15, it may be further deconvoluted into two bands, a gaussian now centered at ~11.6 cm$^{-1}$ and another much broader asymmetric at ~16.9 cm$^{-1}$ with a truncated peak profile in the 1T-3T range, under a gaussian multiconstituent envelop, that may be understood as due to the disruption of individual magnetic sites by the external field perhaps heightened due to the polycrystalline character of our samples. This resonance becomes narrower and comparable to the other magnetic bands increasing fields, all conforming under 7 T a close frequency quartet at 22.8, 21.3, 17.15 and 11.1 cm$^{-1}$ (Fig. 4 (inset)) suggesting that the low fields broadening is consequence of precessing moments under inhomogeneous individual exchange and dampings excluded in the two-site picture. In this regard, it is also worth noting that the ZFC $\omega_{qFM}$ mode at 14.7 cm$^{-1}$ (Fig, S4 in the Supplemental Material [15], upper panel) has a band profile wide enough that may be also reproduced by two bands as for contemplating the existence of a zero field extra weaker intrinsic exchange. This, nonetheless, does not exclude considering the overall behavior of q-AFM- and q-FM-like magnons in LaFeO$_3$ be a suitable standard against the characterization shown in the following subsections for the RFeO$_3$ (R=Pr, Er, Lu) orthoferrites.

### b) PrFeO$_3$

Highly correlated non-Jahn-Teller PrFeO$_3$ is a Fe$^{3+}$ high state distorted orthorhombic perovskite in which the A site rare-earth-oxygen distances embrace eight nearest neighbors out of the cubic twelve, (Fig. 2), conforming a sublattice in which the lanthanide $4f^{\,2}$ shell has CF $^3$H$_4$ ground state which multiplet crystal field splits into nine singlets.

This compound, as for LaFeO$_3$, is in the room temperature $\Gamma_4$ (G$_x$, A$_y$, F$_z$) magnetic representation below the Nèel temperature T$_N$ ~703 K [74, 75]. It also develops below T$_N$ weak ferromagnetism along the <u>c</u> axis due to Fe$^{3+}$ canted moment with angle α(deg) =0.49∘.[69] On decreasing temperature, a broad inflexion in the reciprocal magnetic susceptibility has been associated to a spontaneous spin reorientation turning, between 201 K and 140 K, $\Gamma_4$ ($G_x$, $A_y$, $F_z$) into $\Gamma_2$ ($F_x$, $C_y$, $G_z$) rotating magnetic moments in the _ac_ plane due to 4$f$-3$d$ competing exchanges by $Pr^{3+}$ and $Fe^{3+}$ [21]. However, against this observation neutron diffraction patterns at 8 K confirm the $G_x$ type antiferromagnetic arrangement. It also supports earlier conclusions by which weak Pr-Fe couplings question the possible reorientation phase transition [20].

We will assume that in our measurements the $\Gamma_4$ ($G_x$, $A_y$, $F_z$) representation holds in the temperature range of our studies.

At low temperatures we find a distinctive band at ~16.6 cm$^{-1}$ that becomes sharper and that is also central to two much weaker and broader ones (Fig.3(b)). It softens, undergoing an energy down shift with zero field cooling from 5 K to ~2K, likely consequence of exchange couplings that without an apparent change in the side band profiles retain the high temperature magnetic representation $\Gamma_4$ ($G_x$, $A_y$, $F_z$). The central feature matches inelastic neutron data at 16 cm$^{-1}$ (2. ± 0.1 meV) assigned to the lowest of the 2J+1 $^3H_4$ ground state level, [76. It has an asymmetric band shape reproduced by an overall Weibull profile due to its coalescing into near degeneracy with the bands at ~9.9 cm$^{-1}$ and ~25.5 cm$^{-1}$ contributing to the fused profile (Fig. S6 in the Supplemental Material [15]). These last two are assigned to the $\omega_{qAFM}$ and $\omega_{qFM}$ modes by association with the resonances found in LaFeO$_3$. The emerging picture is close to what we earlier reported for PrCrO$_3$ where it was possible to deconvolute that main band into three components in spite that transition metal anisotropy fields together with the paramagnetic $Pr^{3+}$ moments add random fluctuations in the q-AFM mode dynamically blurring magnetic oscillations [77]. In PrFeO$_3$ this would prevent well defined q-AFM x, y out-of-phase and in-phase oscillations along the _z_ axis and q-FM low frequency cones delineated by the ferromagnetic moment precessing the _z_ axis (Fig. 1).

To have a better picture of the low energy spin dynamics, and possible field induced 3d-4f coupling along spin fluctuations, we also applied an external magnetic field. Fig. S7 in the Supplemental Material [15 shows that now, and as in PrCrO$_3$ [77]**,** the lowest of the 2J+1 $^3H_4$ ground state level behaves as expected for a 4$f$ even electron lanthanide in ground crystal singlets being up to 7T nearly unaffected by the external field [78]**.** There is, however, an effective weakening, broadening, and merging that may mask unresolved field weak dependences. As consequence, we only have significant departures from linearity in the broad feature assimilated to the $\omega_{qFM}$ $Fe^{3+}$ resonant mode. In either case, there is no field induced split degeneracy as found in LaFeO$_3$, but a field increase induced broadening suggesting that magnon splitting may be screened out due to the $Pr^{3+}$ paramagnetism further disturbing magnetic couplings and anisotropies of the in-phase oscillation ($\omega_{qAFM}$) and precession ($\omega_{qFM}$) that in presence of the ground state dipole transition appear merging all excitations into one highly coupled mixed scenario. The relative clean profile for the ground level at weaker fields becomes a band with much less oscillator strength turning near featureless under 7.5 T at 5 K (Fig. 5) This is a likely consequence, again, of magneto dynamical fields in a mixed scenario in which the local microscopically altered lattice structure may also play a role [79].

### c) $LuFeO_3$

The magnetic dependent features found in $LaFeO_3$ are also expected for orthorhombic $P_{bnm}$-$D_{2h}^{16}$ $LuFeO_3$ ($T_N$ ~623 K) [80] since main differences in bulk properties are only introduced by the replacement of La by the also closed electronic shell Lu ($4f^{14}$). Both compounds, $LaFeO_3$ and $LuFeO_3$, share the $\Gamma_4$ ($G_x$, $A_y$, $F_z$) magnetic representation at room temperature [24] as well as canted ferromagnetism here with $Fe^{3+}$ canting angle range $\alpha$(deg.) =0.80∘ [18]- 0.61∘ **[72]** relative to the *a* oriented sublattice. As $Lu^{3+}$ does not carry localized magnetic moment, it does not show neither magnetic compensation nor spin reorientation effects [81]. We find that for both compounds zero field cooled temperature dependences show profiles that may be assigned to magnons. For $LuFeO_3$ they are at $\omega_{qFM}$ ~22.4 cm$^{-1}$ and $\omega_{qAFM}$ ~26.3 cm$^{-1}$ (Fig. 2 (c)) with an extra third band at ~10.4 cm$^{-1}$. To understand this last one it is important to recognize that, in addition to rigid tilting and rotation, the consequence of Lu $4f^{14}$ smaller ionic radius at the A site is the edge sharing inducing ligand distortions that indirectly trigger subtle changes in the octahedral $FeO_6$ B site. The changes in Fe-O bonding due to this ionic size dependence not only affects the known strong antiferromagnetic nearest neighbor Fe-O-Fe exchange interaction but also means a minute but effective local break in the space inversion symmetry allowing $Fe^{3+}$ crystal field $^6A_1$ multiplet transitions. This slight perturbation at the B sites makes possible the detection of CF transitions and solves the puzzle on the appearance and nature of a third absorption in far infrared measurements by Aring and Sievers [82]. It also provides a clue of the lattice role on the origin of the ferroelectric loop in the magnetic ordered phase below ~$T_N$ [83].

The effect introduced by lattice distortions due to smaller lanthanides in RFeO$_3$ has been recognized by phonon Raman scattering measurements [62, 84] and it is most distinctive in changes in far infrared reflectivity when LaFeO$_3$ is compared against LuFeO$_3$.(Figs. S12-S15 in the Supplemental Material [15]) Better defined bands, as for stretching modes around 600 cm$^{-1}$, and the need of extra oscillators in fits (Tables SI-SIV) suggests minute energy differences translated into smaller ion displacements beyond most today's X-ray diffraction resolution and it is at the root of identifying ferroelectric spontaneous polarization in orthorhombic LuFeO$_3$ [85]. It is also close to arguments on lattice strains as its possible origin in RFeO$_3$ thin films making plausible a P*bn2$_1$* space group assignment rather than P*$_{bnm}$*[86] all compounding with the role played by the oxygen electronic polarizability in a ferroelectric phase transition.[87] The oxygen anisotropic volume dependent non-linear polarizability, that in ABO$_3$ introduces a dynamic covalent enhancing in the Oxygen p hybridization with Iron d states [88], allows a closer unified picture for ferroelectricity beyond introducing a foreign antisymmetric coupling of two non-equivalent spin pairs [89]. As it has been stressed many times most ferroelectrics are oxides where coexist displacive and order/disorder dynamics [90].

Fe$^{3+}$ crystal field ground level transitions [91] for low level doped B sites in II- VI compounds are double degenerate and optically active in the lower THz portion of the spectra. They have been associated to transitions of the Fe$^{3+}$ $^6$A$_1$ multiplet in embedded cage in the α-Al$_2$O$_3$ hematite-corundum lattice [92-94] when studying population inversion in Fe$^{3+}$ doped sapphire [95-97]. We associate our local distortion at B sites with these "impurity" induced regimes in which the cubic pure symmetry turns into nearly Oh at distorted sites following the known concurrent structural sequence -cubic (CaTiO$_3$)-orthorhombic (RFeO$_3$)-rhombohedral (α-Fe$_2$O$_3$) [7].

Then the absorption band at ~10 cm$^{-1}$, Fig. 3 (c), and its linear Zeeman branching, (Figs. 6 and S8 in the Supplemental Material [15]) are interpreted as from the sextet Fe$^{3+}$ 3d$^5$ ($^6$S$_{5/2}$) in the octahedral crystal field under a subtle structural distortion [98]. The Fe$^{3+}$ line in LuFeO$_3$ would be the highest Kramers degenerate level of the three expected to split linearly [93-95]. It accounts for electron-electron repulsion, spin-orbit interaction, and crystal-field potentials [46,98-100]. The energy splitting under external magnetic fields allows to estimate the g factor using the linear relation,

$$\Delta E = \mu_B \, g \, B_0 \qquad (5)$$

that holds in the lower field regime. $\Delta E$ is the energy level Zeeman split shown in Fig. 6 (inset), $\mu_B$ denotes the Bohr magneton ($\mu_B$=5.788 10$^{-5}$ eV/T); g is a dimensionless constant, and $B_0$ the applied external magnetic field [101]. Our data at 5 K yields $g_{net}$=1.986 that, taking it as result from a close to four near inequivalent iron sites, coincides within the experimental error with $g_e$ = 2.0023 for the free electron. It is the about value expected for high spin $d^5$ configuration as in transition metals like $Fe^{3+}$, meaning a very small orbital contribution to the magnetic moment (quenched orbital momentum) with negligible spin orbit coupling.

The lattice change may be also considered an intermediate step before further departing toward consolidating the hexagonal non-centrosymmetric non-perovskite arrangement. This is a metastable hexagonal structural configuration described by the polar P63cm space group, as in H-$LuFeO_3$, where a trivalent transition metal ion occupies a trigonal bipyramidal site ($RO_5$) [102, 103] rendering the view that low level A-site substitutions in $ABO_3$ compounds are functional to weak ferroelectricity [104].

As noted, at slightly higher frequencies than the $Fe^{3+}$ ground transition the two-zone center resonant magnon modes, $\omega_{qFM}$ and $\omega_{qAFM}$, are found in zero field cooled runs at ~22.4 cm$^{-1}$ and ~26.3 cm$^{-1}$ respectively. The peak positions exhibit a behavior analogous to those in $LaFeO_3$ and are align with an earlier peak quotation by Aring and Sievers [82]. However, in $LuFeO_3$ both resonances under applied fields (depicted in the peripheral insets in Fig. S9 in the Supplemental Material [15]) seem linked by a strong "bridge" that it is only reproduced in fits by introducing an extra background band with distinctive increasing asymmetry. They evenly diverge out the center frame under increasing fields. Herrman [24] pointed out that when considering the four $Fe^{3+}$ magnetic sites, rather than the two sublattice model, it is necessary to take into account not only the so called plain "overt" canting but also a hidden canting mechanism producing no net magnetization. It was found that this last mechanism is particularly significant for the case when the antisymmetric exchange is smaller than the anisotropy energy (A>>D), a case suggested by our $RFeO_3$ (R=rare earth) measurements. We may then conjecture that the optical activity of the bridging excitation at frequencies between the FM and AFM modes stems from an unresolved constituent of hidden canting inducing coupling between exchange and magnon resonances. The exchange resonances by themselves are expected to be optically inactive in antiferromagnetics.[24]

We also find a quasiundulating pattern at ~15 cm$^{-1}$ (see fig. S9 in the Supplemental Material [15]. This pattern consists of four emerging very weak maxima indicated by the red arrow in the upper-left 1T inset. At these intermediate frequencies, where the Fe$^{3+}$ branching intersects with the diverging q-FM magnon, (as also shown in fig. 6), zooming on the 15 cm$^{-1}$ to 20 cm$^{-1}$ window allows to deconvolute each weak peak by partially superimposing two broad Gaussians. For simplicity, the remaining three are omitted. We interpret the four individual peaks as arising from inequivalent magnetic sites that ought to be taken into account if considering inequivalent four Fe$^{3+}$ magnetic sites rather than the two sublattices of the present working hypothesis [24]. These weak features also appear with similar frequency spread on the mode higher-frequency branching of ωqAFM (see also Fig. S10 in the Supplemental Material [15]) corroborating that both resonances, ωqAFM and ωqFM, are indeed intertwined in a collective complex trend. This collective behavior, along with the effects of polycrystallinity that blurs individual phase motion, may be then associated to the rare earth-induced A site lattice perturbations. These perturbations, at the root of spin canting single-ion anisotropy [18], is the cause the optical activity of the Fe$^{3+}$ ion in disturbed lattice sites as addressed in previous and following paragraphs.

### d) ErFeO$_3$

ErFeO$_3$ (T$_N$ ~633 K) [31, 63] orthorhombic P$_{bnm}$-D$_{2h}^{16}$ is perhaps one the most study compound of the ferrite family [105]. Starting at about 220 K and on cooling [106] Er$^{3+}$ moments start forcing the prevailing magnetic oriented iron moments in the Γ$_4$ (G$_x$, F$_z$) representation to rotate in 90° into the F$_2$ (G$_z$, F$_x$) *a*-axis alignment in an interval from T$_1$= 87 K (90 K) to T$_2$=96.6 K (103 K). [107-112]. This mostly magnetic displacive phase transition, in which has also been reported strong spin-lattice coupling, is common to canted ferromagnetism in all partially filled rare earth *f*-odd shell compounds [8, 30,31, 99,102].

We studied in this Γ$_2$ (G$_z$, F$_x$) phase the Fe$^{3+}$ ($^6$A$_1$) and the Er$^{3+}$ ($^4$I$_{15/2}$) manifolds as well as the two spin resonances. As it is shown in the mapping of an incremental field plot, Fig. 7, the Fe$^{3+}$($^6$A$_1$) transition, extrapolated in our detection limit to ~4 cm$^{-1}$, undergoes a linear Zeeman split in a perturbed A site close behaving the discussed in the section for LuFeO$_3$. However, unlike the symmetric profile in LuFeO$_3$ (Fig. 6), it is now biased toward higher energies due to a drift associated to the Er$^{3+}$-Fe$^{3+}$ magnetic exchange.

Fig. 8 (a) shows that on cooling the intensity of the q-AFM resonance increases continuously below the reorientation temperature $T_{RS}$ peaking at about 5 K to then turning into a sharp decrease as $ErFeO_3$ gets closer to the $\Gamma_1$ ($A_x$, $G_y$, $C_z$) phase. In this phase, after softening, it merges with the much weaker hardening ferromagnetic mode ($\omega_{qFM}$). This results in a single much broad feature characteristic of $\Gamma_1$ phase as the $Fe^{3+}$ spins undergo a gradual $G_z$ to $G_y$ rotation due $Er^{3+}$ exchange coupling leading to cooperative antiferromagnetic for $Er^{3+}$ below $T_N(Er)$ ~4.5 K. (Fig. 8 (a, b)) This behavior is similar to what has been reported for $ErCrO_3$ in the same temperature range [77]. The sharp decrease of the $\omega_{qAFM}$ mode intensity signals the onset of the $G_{xy}F_x$ mixed spin arrangement [113] at the time that corresponding magnetic structure tur ns into fully $C_z$ antiferromagnetic oriented for $Er^{3+}$ and magnetic $C_yG_zF_x$ for $Fe^{3+}$ [114,115] in a potential low temperature monoclinic space group $P_{2_1/m}$. It s also worth to underline that in our studies we do not observe an anomaly, electric dipole related, that might be assigned at the lowest temperatures to an electromagnon even upon applying moderate external magnetic field as it was reported in nominally isomorph $TbFeO_3$ [116] and $DyFeO_3$ [117]. In an electromagnon, rare-earth magnetic moments would align with the field (either from perturbating $Fe^{3+}$ or external) causing a change in the electric polarization that would be possible detected as extra absorption in THz-far infrared measurements.

The q-FM and q-AFM resonances in $ErFeO_3$ at $\omega_{qFM}$~21.5 $cm^{-1}$ and $\omega_{qAFM}$~ 31.5 $cm^{-1}$ have an equally weak field dependence up to the compensation temperature $T_{CMP}$ [113]. This is the temperature at which the antiferromagnetic exchange interaction of paramagnetic $Er^{3+}$ moment equals the near constant moment of $Fe^{3+}$ spins meaning a net polarization of the $Er^{3+}$ spins antiparallel to the canted $Fe^{3+}$ [118, 119]. That is, the negative exchange interaction of the iron and erbium yields a total spontaneous moment written as

$$M = M_0 + \varkappa_R \cdot H_0 \qquad (6)$$

being $M_0$ the ferromagnetic moment, $\varkappa_R$ the paramagnetic rare earth susceptibility, and $H_0$ the exchange field at the rare earth by the irons. With negative exchange interaction the rare earth induced magnetic moment decreases and so the total magnetic moment diminishes up to vanishing at $T_{CMP}$ [120]. A vanishing point also remains upon applying an external magnetic field such that at temperatures above it the canted ferromagnetism is directed along the field.

$ErFeO_3$ field dependent induced absorptions from THz resonances and Zeeman splits are shown using as reference the ZFC spectrum at 5 K in Fig. 9. As it was in their temperature dependence, the induced change at the q-AFM resonance undergoes a huge enhancement while the rather weak q-FM mode behaves accordingly to the previous cases. However, at difference with what we found for the other compounds they do not diverge split upon applying up to 7 T. It appears as if their divergency is compensated by the $Fe^{3+}$-$Er^{3+}$ exchange yielding a very narrow picture that, unsolved at our working resolution, increases in intensity as assuming superposing two gaussians.

Zero field ratios calculated using the spectra taken for the initial and after the 7 T run yield extra absorptions at both magnons frequencies bringing up the delicate nature of the exchange on these modes. Interpreted as a memory effect this behavior may also help to explain the origin of the broad and intriguing extra line reported by Mikhaylovskiy et al [121]. Our spectra show that at zero field the $Fe^{3+}$ unfolded multiplet and zone center magnons are separated by at least 15 cm$^{-1}$ thus ruling out a zone center interference as a possible explanation. Rather, the unexplained extra band at 11.7 cm$^{-1}$ (0.35 THz) shouldering the ferromagnetic mode might have been created by a laser induced distortion changing locally the effective temperature within the dynamics of the resonant mode.

Alongside the temperature dependent-field dependent changes of the magnetic resonances, the overall picture for the $Fe^{3+}$ and $Er^{3+}$ multiplet transitions remain essentially unaltered (Fig. S11 in the Supplemental Material [15]). We note, however, that bands of the Zeeman $Fe^{3+}$ lower energy branch associated to intensity increments, at increasing field ratios, correspond to a shallow response in the branch facing the $Er^{3+}$ ($^4I_{15/2}$) Zeeman split at higher frequency. Only when ratios for this last one are calculated decreasing fields, we retrieve meaningful gaussian profiles (blue traces in Figs.10 and S11 in the Supplemental Material [15]) suggesting depopulated levels due to $Er^{3+}$ exchange. The applied field disrupts selectively the known symmetric part of the exchange energy between the two $Fe^{3+}$ magnetic sublattices $S_i$ and $S_j$ and thus the zero field superexchange involving nearest neighbors Fe+ ions via intermediate $O^{-2}$ ion. In zero field superexchange one spin up electron virtually hops to oxygen forming an up-down pair and back to the same spin orientation [122]. We find that this mechanism is particularly sensible to the paramagnetic $Er^{3+}$ lowest Zeeman branch when the higher $Fe^{3+}$ and lower $Er^{3+}$, branches meet at $T_{CMP}$ ~40 K at about 30 cm$^{-1}$ and ~5 T (solid arrow in Fig. 5). In this scenario, and in contrast with the clean picture suggested by the LuFeO$_3$ analysis, the calculation of the g factor now reflects increasing couplings and anisotropies induced by the ligand field in what would imply a net effective result. Choosing energy levels at intermediate fields from the skew Zeeman split shown in fig. 10(a) and replacing $B_0$ by $B_{eff} = (B_0 + L_{ocal})$ and g by $g_{eff}$ in eq (5) [102] we found $L_{ocal}$ ~1.08 T that in turn results in a ballpark $g_{eff}$ ~ $10.0 \pm 0.5$.

While there is Fe$^{3+}$ ($^6A_1$) population inversion at all temperatures, in the Er$^{3+}$ branch it is only associated to paramagnetic Er$^{3+}$ random dipoles. In ErFeO$_3$ the field dependent extra absorption at 40 K about 30 cm$^{-1}$ and ~5 T diminishes as Er$^{3+}$ gets closer to its ordered antiferromagnetic phase where branching becomes near field independent. At 5 K the induced change becomes undetectable and at 2.4 K has totally disappeared allowing undistorted Gaussian profiles in the Fe$^{3+}$ ($^6A_1$) higher frequency branch (Figs.10(b) and S11 in the Supplemental Material [15]).

## e) Er$^{3+}$ multiplet and phonon field dependences

To get a full view of the temperature field dependence of the Er$^{3+}$ ($^4I_{15/2}$) multiplet it is necessary to run far infrared spectra. In this range the near normal reflectivity of ErFeO$_3$ at 5K, Fig. S13 in the Supplemental Material [15], is dominated by phonon profiles that may be broadly grouped into a lower frequency region centered at ~ 200 cm$^{-1}$ for lattice vibrations in which the rare earth moves against FeO$_6$ sublattice, a second one centered ~350 cm$^{-1}$ that it is characterized by Fe scissor and stretching modes as well as octahedral librations, and a third one for oxygen breathing-stretching-modes at ~600 cm$^{-1}$. When this reflectivity is matched to the absorption ratio of 1 T against 0 T spectra, Fig. 11 (a), it is straight forward to locate the Er$^{3+}$ ($4f^{11}$) three Kramers ($^4I_{15/2}$) doublets as their magnetic signature appears at 49.5 cm$^{-1}$ (6.14 meV), 110.5 cm$^{-1}$ (13.7 meV) and 167.3 cm$^{-1}$ (20.74 meV). They are in agreement with recent inelastic neutron scattering measurements [115, 120], and earlier estimates [123].

Just 0.5 T is enough to lift the Er$^{3+}$ Kramer degeneracy by ~5 cm$^{-1}$ (0.62 meV) into two weaker bands. At temperatures below compensation point $T_{cmp}$~40 K, and increasing the fields, Fig. 12, they turn into strong asymmetric Zeeman splits correlated with the Fe$^{3+}$ branching at lower frequencies. While this asymmetric picture is seen for the three transitions, it is particularly clear for the third level at 167.3 cm$^{-1}$ being symmetric above $T_{cmp}$ to then have a field insensitive lower frequency branch as it becomes associated to the $Fe^{3+}$ exchange at and below $T_{cmp}$ (Fig. S16 in the Supplemental Material [15]).

$Er^{3+}$ levels at 110.5 $cm^{-1}$ and 167.3 $cm^{-1}$ fall at known reflectivity phonon frequencies for vibrational modes involving rare-earth displacements and, least for one case, seems to be also related to an $A_g$ phonon in Raman spectra that for $LaFeO_3$ is at 84.5 $cm^{-1}$[50]. Fig. 11 (a) also shows a dotted line across a continuous envelop in the field induced absorption of the 1 T/ 0 T ratio coinciding with the region of external lattice vibrations. It suggests localized charge delocalization for the frequencies where phonons are found interacting with the crystal field levels and thus spins. This is at the root of the so-called 'phonomagnetism" where phonons induce magnetism by light excited lattice distortion modifying the rare earth-transition exchange interaction. It has been recently shown that non-thermal resonant driven by light phonons may be be use to manipulate magnetic states in exchange interactions between rare earth orbitals and transition metal spins switching antiferromagnetic and weakly ferromagnetic spin orders. [124]

Next, at increasing frequencies, there are three individual field enhancements centered at ~400 $cm^{-1}$, the frequency for $Fe^{3+}$ torsional vibrations (vertical arrows in Fig. 11(a)). Each phonon corresponds to a field induced absorption that becomes stronger at frequencies closer to the respective longitudinal optical mode associated to the reflectivity transverse optical-longitudinal optical (TO-LO) reststrahlen split. Non-linear vibrational modes are also found to drive rotations and displacements mimicking the application of a magnetic fields yielding spin precession as a tool for stimulation of novel opto-magnetic phenomena. [125]

Also shown in Fig. S13, in the Supplemental Material [15], that under 1T we do not find a magnetic response for breathing stretching modes centered at ~600 $cm^{-1}$ where only oxygens move. Field dependences are only detected for vibrations involving moving magnetic ions, in our case $Er^{3+}$ and $Fe^{3+}$, fulfilling the basic premise in TO-LO magnetoelectrics by which ion displacements couple electric and magnetic contributions appearing in the macroscopic field [9]. This contrasts with dielectric insulators where macroscopic field linked to a longitudinal optical mode is only associated to long-range electric fields as restoring force due to Coulomb interactions and by which LO frequency is a minimum over its reststrahlen [10].

## Conclusions

Summarizing, we studied low temperature-low energy absorptions of LaFeO$_3$ antiferromagnetic and ferromagnetic magnons at q$\omega_{FM}$ ~26.7 cm$^{-1}$ and q$\omega_{AFM}$ ~31.4 cm$^{-1}$ in the $\Gamma_4$ (G$_x$, A$_y$, F$_z$) representation. We found that their degeneracy is lifted about linearly by an applied fields up to 7 T. In isomorph LuFeO$_3$ in addition to these two spin modes, now peaking at q$\omega_{FM}$ ~22.4 cm$^{-1}$ and q$\omega_{AMF}$ ~26.3 cm$^{-1}$, there is a Fe$^{3+}$ crystal field $^6$A$_1$ transition at ~10.4 cm$^{-1}$ which activity is tied to ligand distortions induced by Lu$^{3+}$ 4$f^{14}$ smaller ionic radius. This A site deformation triggers subtle lattice changes at the perovskite B site and allows a linear Zeeman split up to 7 T. It is also found in ErFeO$_3$ (Kramers 4$f^{11}$ Er$^{3+}$; $\Gamma_2$ (F$_x$, C$_y$, G$_z$) <T$_{SR}$ ~93 K), where now the Fe$^{3+}$ Zeeman branching is strongly biased toward higher energies due to 3d-4f exchange. Within our 0.5 cm$^{-1}$ working resolution at 5 K, and in contrast with magnons in RFeO$_3$ (R=La, Lu), there is no field induced splits but a low temperature 13-fold increase in the intensity of the resonance antiferro/ferro mode ratio. It is also remarkable the CF matching population balance between Fe$^{3+}$ higher and Er$^{3+}$ lower Zeeman-split branches.

Er$^{3+}$ ($^4$I$_{15/2}$) transition energies from the ground multiplet at 110.5 cm$^{-1}$ and 167.3 cm$^{-1}$ coincide with external lattice mode frequencies suggesting strong crystal field-phonon couplings in a scenario for lattice driven spin-phonon interactions. Absorption spectra ratios under mild external fields reveal a magnetic dependent local quasicontinuum in the Er-O vibrational region and enhancement in longitudinal optical mode macroscopic fields associated to Fe$^{3+}$ torsional modes, i.e., field dependences are only detected for vibrations involving moving magnetic ions.

Antiferro- and ferro- resonances turn in PrFeO$_3$ much broader when non-Kramers Pr$^{3+}$, with two unpaired 4$f^2$ electrons, introduces ligand changes at the A site leading the q$\omega_{AFM}$ mode and the lowest Pr$^{3+}$ CF transition into near degeneracy. This and the q$\omega_{FM}$ merge at 7 T into a single broad mostly unresolved feature.

We conclude that low energy excitations in RFeO$_3$ (R=rare earth) strongly depend on the lanthanide ionic size thus indivisibly tied to the mechanism associated to the origin of canted ferromagnetism. Minute lattice displacements in the perovskite basic BO$_6$ and AO$_{12}$ building blocks also underlies the fact that the ferroelectric instabilities might already be present above Tc at sites with thermal distribution of the ionic motion. Peaking at actual centrosymmetric sites may really be due to some multi-well configuration as most ferroelectrics appear to be in the middle ground represented by shallow double wells and anharmonic rough single wells.[126] This argument motivates the possibility of assigning non-centrosymmetry to those distorted perovskites with smaller rare earth ionic radius as for the a P*bn2₁* space group. I.e., with $Fe^{3+}$ not at a center of inversion in a rigid ion approximation. Then, it is necessary to also take into account the hybridization between oxygen p and transition-metal d electrons. As a feature in driving the ferroelectric instability, it will dynamically enhance the oxygen volume dependent electronic polarizability that it is tied to the emergence of the weak spontaneous polarization within the framework of most known ferroelectric oxides [90, 126].

# Acknowledgements

The authors are pleased to acknowledge the enlightening insights and full support at the THz beamline by K. Holldack (Department Optics & Beamlines, Helmholtz-Zentrum für Materialien und Energie GmbH (HZB), D-12489) during the development of the present work. NEM is indebted to BESSYII at the Helmholtz-Zentrum Berlin für Materialien und Energie for beamtime allocation under proposals 201-09204-ST, 202-09707-ST, 212-10296-ST, 221-10915-ST, 221-10941-ST, 222-11422-ST and 222-11426-ST and for financial assistance supported by the project CALIPSO plus under the Grant Agreement 730872 from the European Union Framework Program for Research and Innovation HORIZON 2020. He also thanks the laboratory on Conditions Extrêmes et Matériaux: Haute Température et Irradiation - UPR3079 CNRS (C.E.M.H.T.I.)) in Orléans and the Groupement de Recherche Matériaux Microélectronique Acoustique Nanotechnologies (GREMAN- UMR 7347 CNRS)- Université de Tours, for sharing expertise on research and financial support performing far infrared reflectivity measurements. JAA acknowledges the ILL-Grenoble for the allowed neutron time, and the financial support of the Spanish Ministry for Science and Innovation (MCIN/AEI/10.13039/501100011033) for granting the project number: PID2021-122477OB-I00-R.

# Figure Captions

**Figure 1.** (color online) Low energy resonant modes for antiferromagnets: (a) higher energy quasiantiferromagnetic ($\omega_{qAFM}$) resonance built on x, y out-of-phase oscillations and in-phase oscillations along the $\underline{z}$ axis; (b) quasiferromagnetic ($\omega_{qFM}$) resonance named after the distorted low frequency cone defined by the ferromagnetic moment precessing the $\underline{z}$ axis. After Hernan [24] and Constable et al [35].

**Figure 2.** (color online) Change of rare earth coordination number in the orthorhombic perovskite polyhedral A site as the lanthanide ionic radius decreases from $La^{3+}$ to $Lu^{3+}$ distorting the higher temperature cubic phase $AO_{12}$ cage [11, 44, 45, 47, 48].

**Figure 3.** (color online) (a) Zone center temperature quasiantiferromagnetic ($\omega_{AqAFM}$) and quasiferromagnetic ($\omega_{qFM}$) magnon modes of $LaFeO_3$ using the 110 K spectrum as normalizing reference; (b) Zone center temperature dependent first excited state of the $Pr^{3+}(4f^{\,2})$ $^3H_4$ manifold and weaker side bands assigned to the $\omega_{qAFM}$ and $\omega_{qFM}$ resonances in $PrFeO_3$ using the 110 K spectrum as normalizing reference. Inset; details showing the undergoing peak softening in the 5 K to ~2.0 K interval. (c) Zone center temperature dependence of the band associated to defect site $Fe^{3+}$ and quasiantiferromagnetic ($\omega_{qAFM}$) and quasiferromagnetic ($\omega_{qFM}$) magnon modes of $LuFeO_3$ using the 110 K spectrum as normalizing reference.

**Figure 4.** (color online) $LaFeO_3$ quasiantiferromagnetic and quasiferromagnetic spin wave resonances as function of the applied magnetic field $B_0$ at 5 K. Inset: peak positions of the shown absorptions after gaussian fits (Fig. S4 in the supplemental material[15])).

**Figure 5.** (color online) First excited state of the $Pr^{3+}(4f^{\,2})$ $^3H_4$ manifold and weaker side bands assigned to the $\omega_{qFM}$ and $\omega_{qAFM}$ resonances in $PrFeO_3$ as function of the applied magnetic field $B_0$ at 5 K. Inset: Inset: peak positions of the shown absorptions after gaussian and Weibull fits (Fig. S7 in the supplemental material[15])).

**Figure 6**. Zone center $Fe^{3+}$ ($^6A_1$) crystal field transition and quasiantiferromagnetic $\omega_{qAFM}$ and quasiferromagnetic $\omega_{qFM}$ magnon modes of $LuFeO_3$ as function of the applied magnetic field $B_0$ at 5 K. Inset: peak positions of the shown absorptions after gaussian and Weibull fits (Fig. S9 in the supplemental material[15])).

**Figure 7** (color online) $ErFeO_3$ sequential absorption ratios $(B_{0j}+0.5T)/B_{0j}$, $B_{0j}$ is the applied field at the $j^{th}$ incremental step, projecting field dependent Zeeman splits of $Fe^{3+}$ ($^6A_1$) and $Er^{3+}$ ($^4I_{15/2}$) and resonant magnon mode (dashed arrows) at the compensation temperature ~ 40 K. Note that both resonances are inside a triangle defined by the highest $Fe^{3+}$ ($^6A_1$) and the lowest $Er^{3+}$ ($^4I_{15/2}$) branch exchange converging at ~5 T. The solid arrow points to the anomaly associated to paramagnetic $Er^{3+}$.

**Figure 8.** (color online) (a) $ErFeO_3$ temperature dependent of the resonance ratios using the 108 K spectrum as reference, inset: same data vertically displaced showing the convergence of both magnons toward the onset of the lower temperature $\Gamma_1$ representation;(b) Applied field dependent $ErFeO_3$ magnon absorption in the ZFC $\Gamma_1$ ($A_x$, $G_y$, $C_z$) phase, at 2.4 K, below and at the interfering $Fe^{3+}$ Zeeman field split.

**Figure 9.** (color online) $ErFeO_3$ absorption ratios $(B_{0J}/0.0T)$, $B_{0j}$ is the applied field at the $j^{th}$ incremental step, for the THz excitations and exchange biased zone center $Fe^{3+}$($^6A_1$) Zeeman split at 5 K. The full line in the back panel shows memory effects brought up by calculating the ratio of spectra of measurements done at the initial and end run at zero field and 5 K.

**Figure 10** (color online) (a) applied field dependent Zeeman split of $Fe^{3+}$ ($^6A_1$) and $\Gamma_2$ magnons, left axis: sequential absorption ratios $(B_{0j}+0.5T)/B_{0j}$, $B_{0j}$ is the applied field at the $j^{th}$ incremental step at 5 K, right axis: sequential absorption ratios $B_{0j}/(B_{0j}+0.5T)$ calculated decreasing fields at 5 K.(b) Applied field dependent Zeeman split of $Fe^{3+}$ ($^6A_1$) , left axis: sequential absorption ratios $(B_{0j}+0.5T)/B_{0j}$, $B_{0j}$ is the applied field at the $j^{th}$ incremental step at 2.4 K, right axis: sequential absorption ratios $B_{0j}/(B_{0j}+0.5T)$ calculated decreasing fields at 2.4 K. The measured profile for the $\Gamma_1$ magnon is shown centered at ~28.5 cm$^{-1}$.

**Figure 11.** (color online) ErFeO$_3$ far infrared absorption, reflectivity, and 1.0 T/0.0 T ZFC absorption ratio at 5 K.[15] (a) Dashed lines: ErFeO$_3$ powder embedded polyethylene pellet absorption spectra at 0 T and 1 T; circle and full line: polycrystal reflectivity spectra and multioscillator fit respectively; full lines: 1.0 T/0.0 T ZFC absorption ratio; asterisks: peak position for the 49.5 cm$^{-1}$ (6.14 meV), 110.5 cm$^{-1}$ (13.7 meV) and 167.3 cm$^{-1}$ (20.74 meV) transitions of the Er$^{3+}$ ($^4$I$_{15/2}$) manifold; (b) vertically displaced reflectivity of RFeO$_3$. (R=La, triangle; Er, square; Lu, diamond) showing relative changes in the R$^{3+}$-O vibrational profiles induced by the change in the lanthanide size at 80 K. Asterisks: ZCF transitions of the Er$^{3+}$ ($^4$I$_{15/2}$) manifold associated to lattice phonons at the respective frequencies.

**Figure 12.** (color online) (a) Ratio of ErFeO$_3$ powder embedded polyethylene pellet 5 K spectra at 1 T and 0 T absorptions peaking at the 49.5 cm$^{-1}$ (6.14 meV), 110.5 cm$^{-1}$ (13.7 meV), and 167.3 cm$^{-1}$ (20.74 meV) transitions of the Er$^{3+}$ ($^4$I$_{15/2}$) manifold; (b), (c), (d) vertically offset Zeeman split sequential absorption ratios (B$_{0j}$+0.5T)/B$_{0j}$, B$_{0j}$ is the applied field at the j$^{th}$ incremental step, of the Er$^{3+}$ ($^4$I$_{15/2}$) levels showing the bias induced by the Fe$^{3+}$ exchange on the lower frequency branch (see also Fig.S16 in the supplemental material[15])).

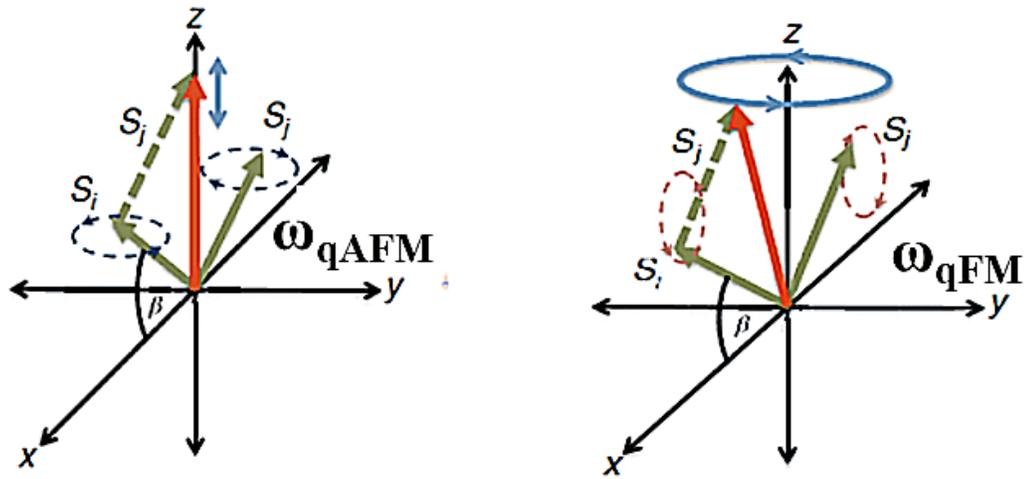

(a) Quasi-antiferromagnetic spin mode  (b) Quasi-ferromagnetic spin mode

Fig. 1
MASSA et al.

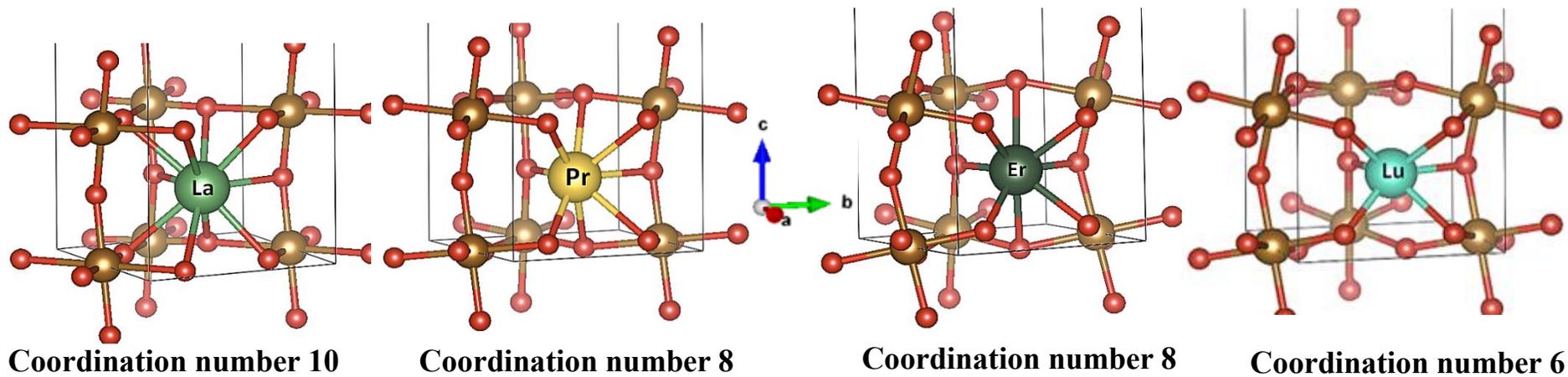

**Coordination number 10**  **Coordination number 8**  **Coordination number 8**  **Coordination number 6**

**Fig. 2**
**MASSA et al**

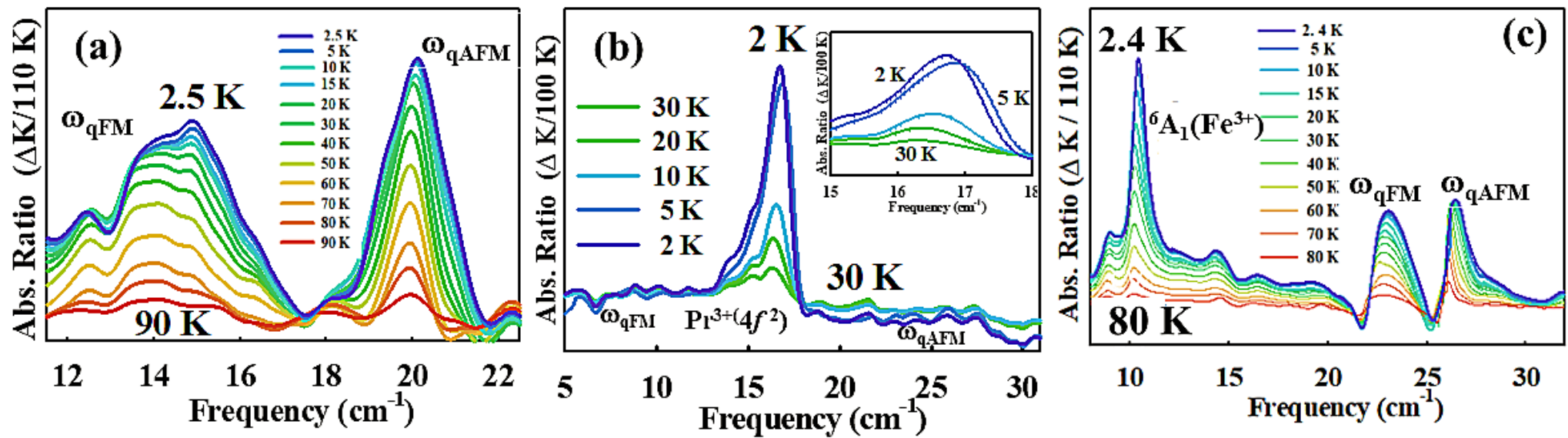

Fig. 3
MASSA et al

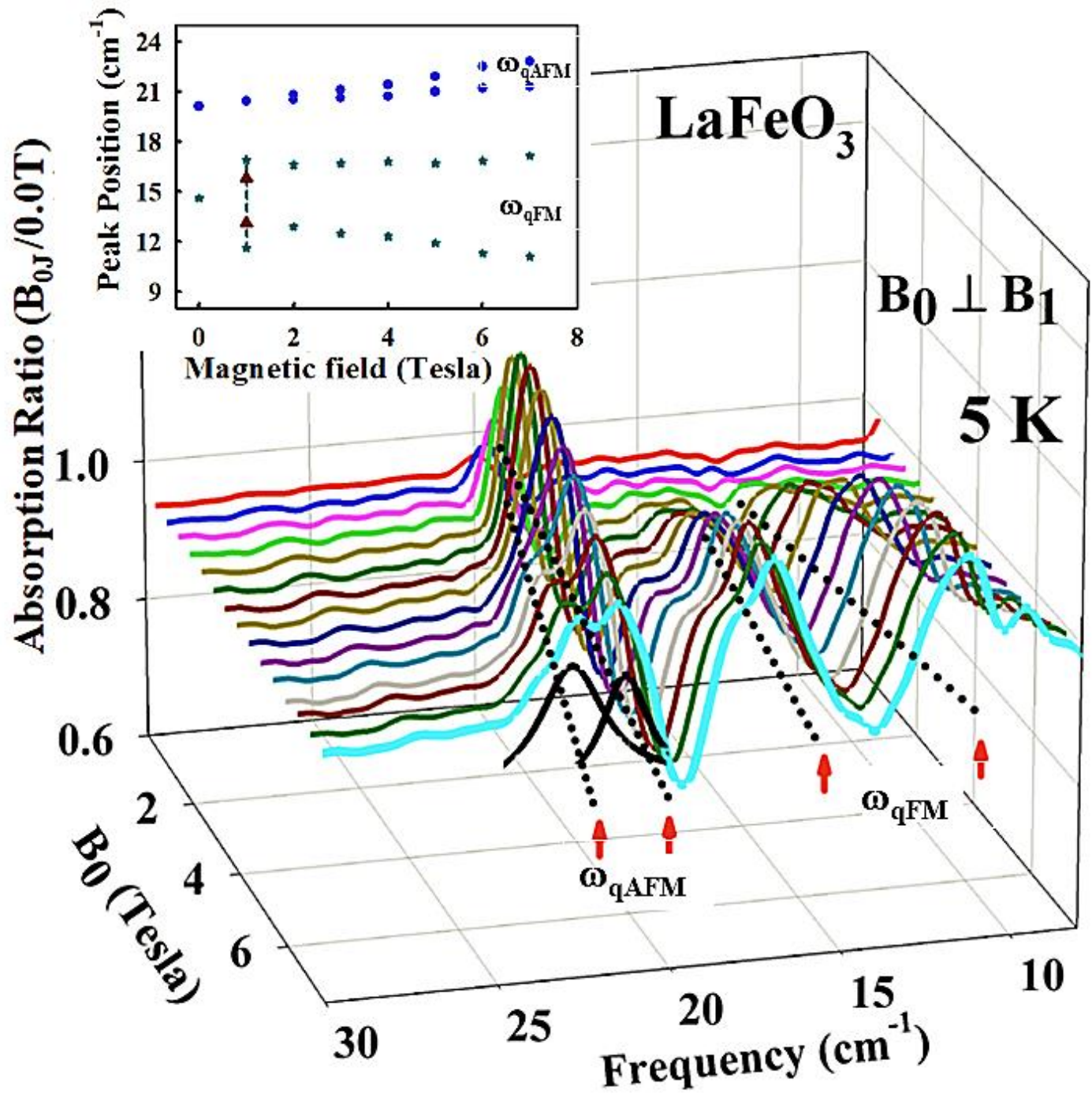

Fig. 4 MASSA et al

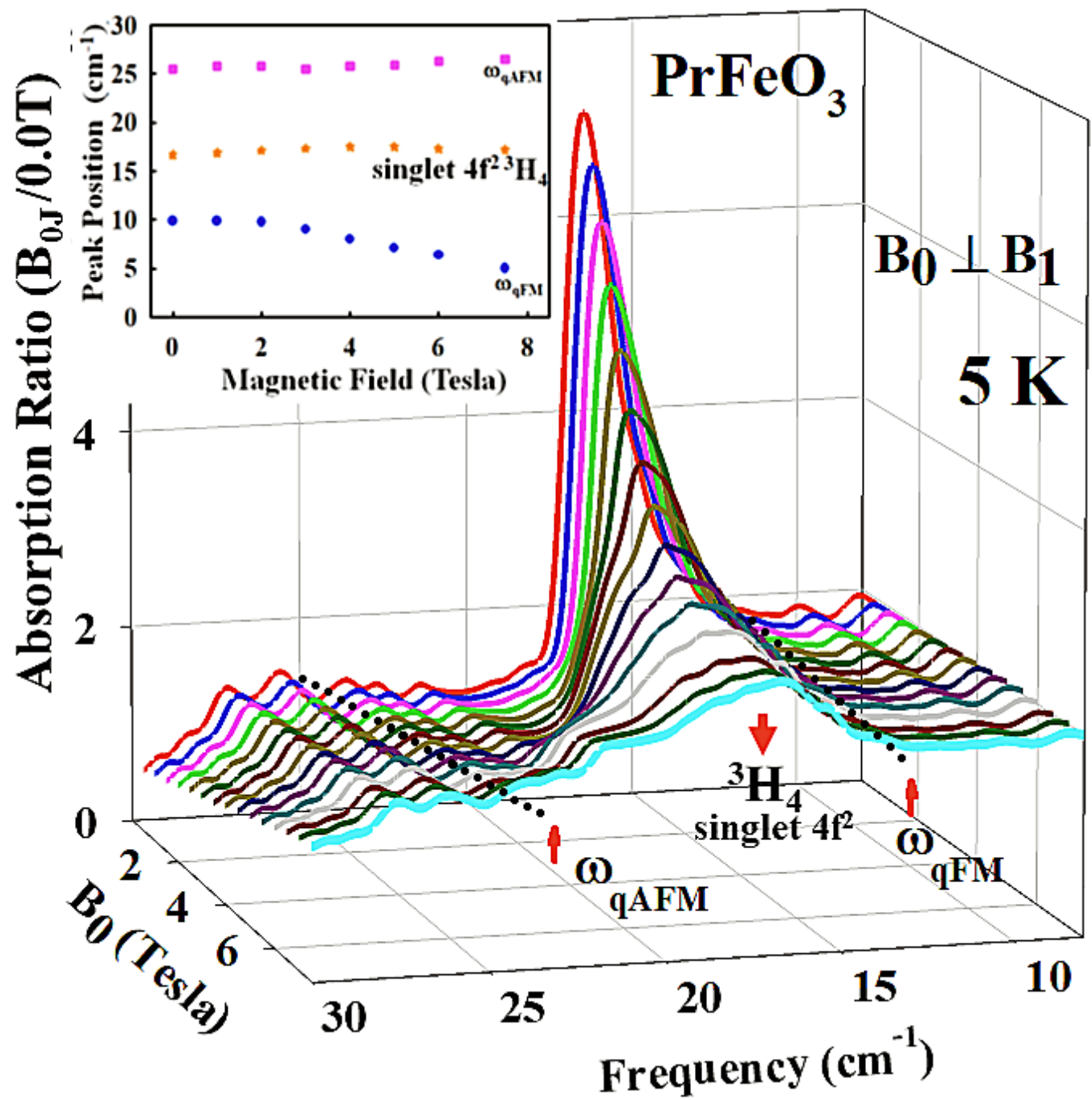

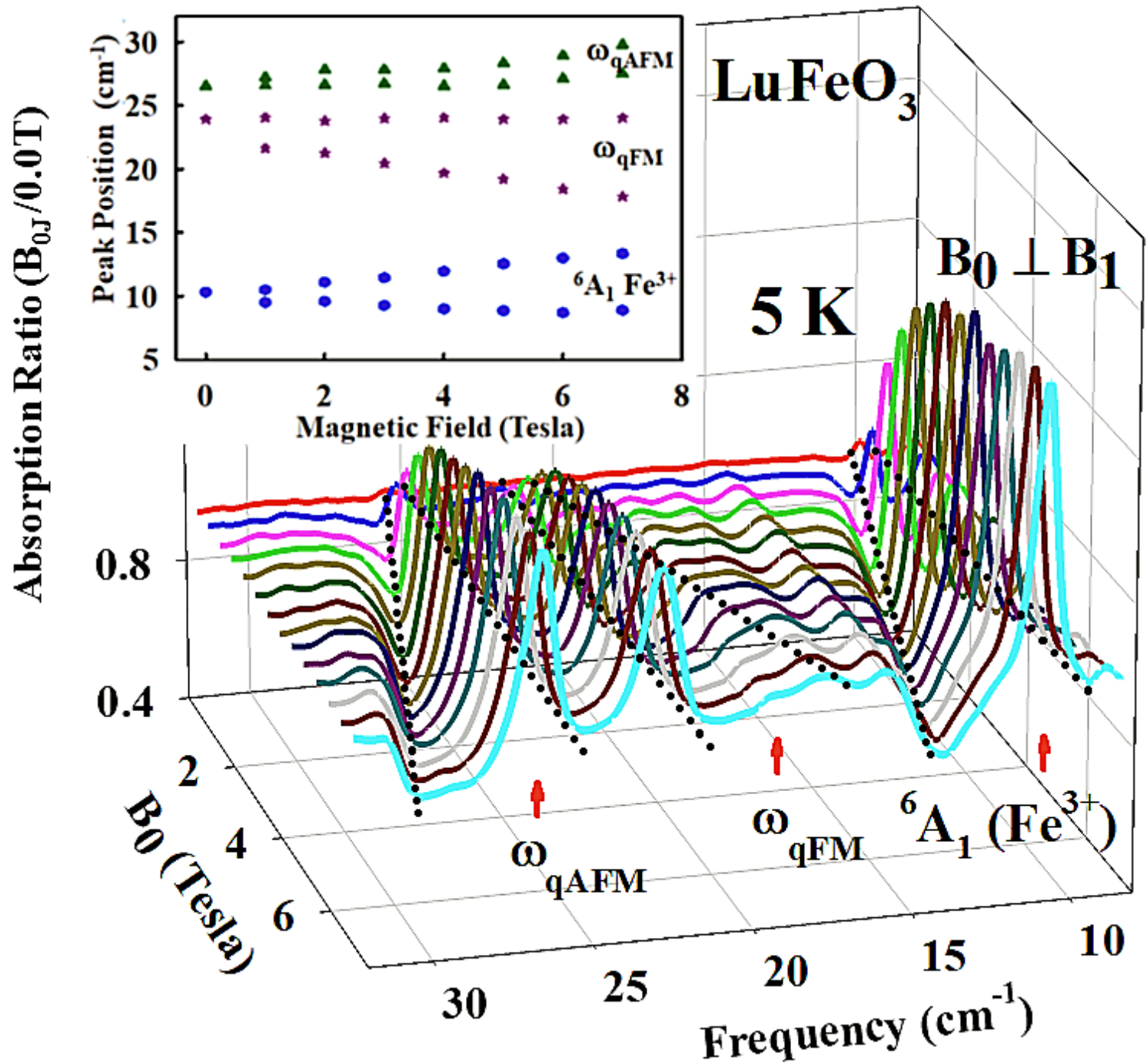

Fig.6 Massa et al

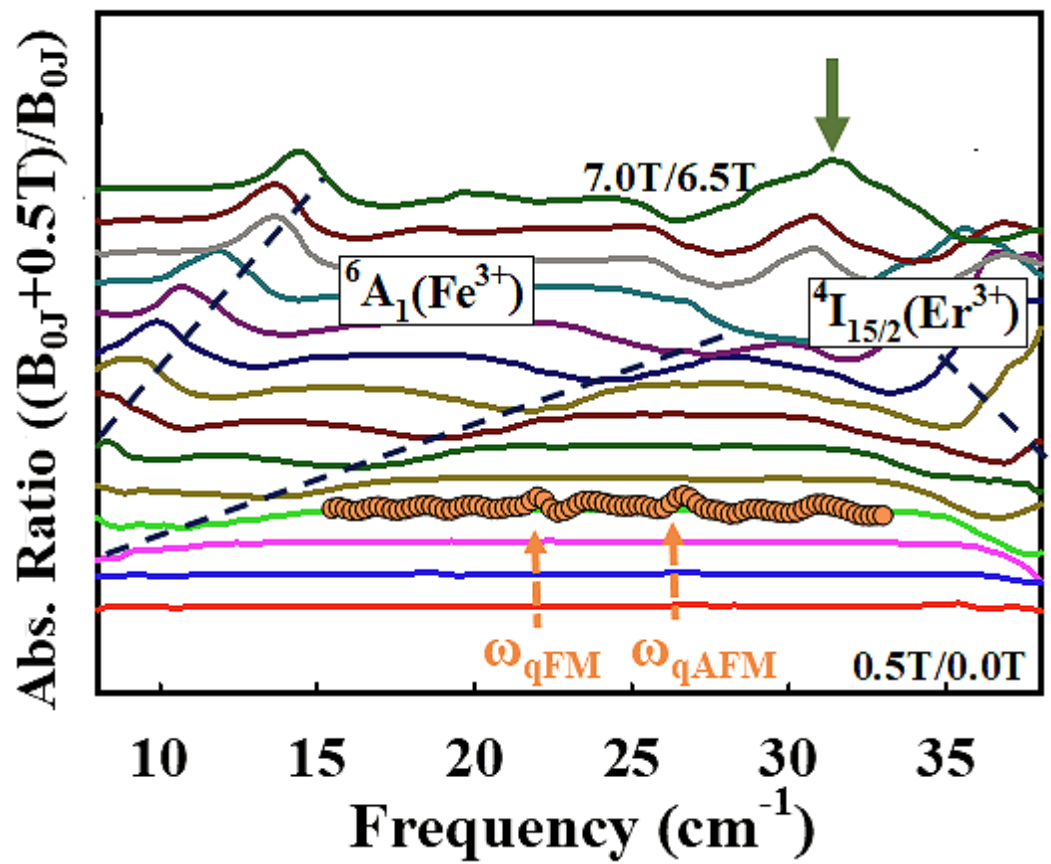

Fig. 7
MASSA et al

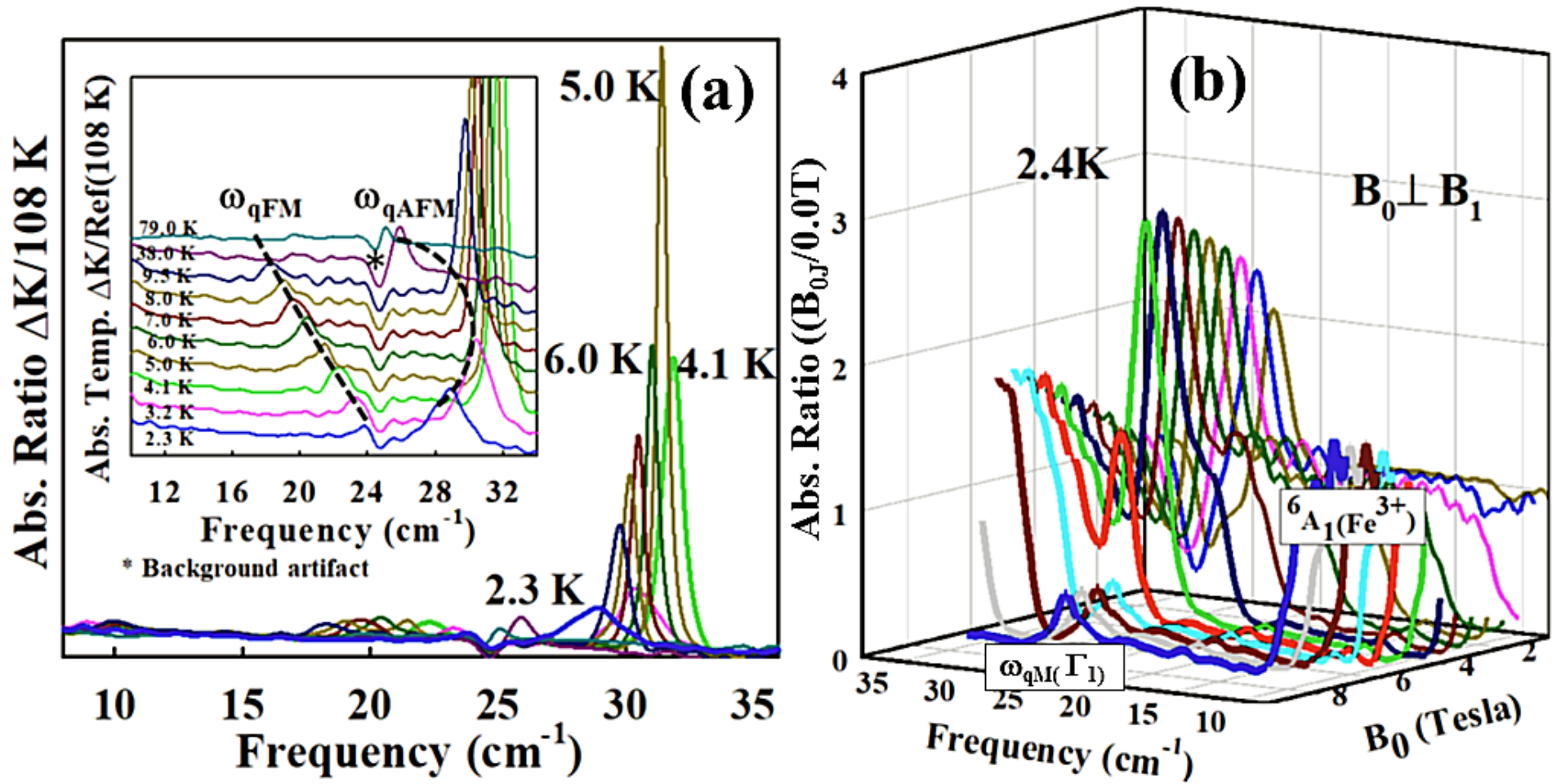

Fig. 8
MASSA et al

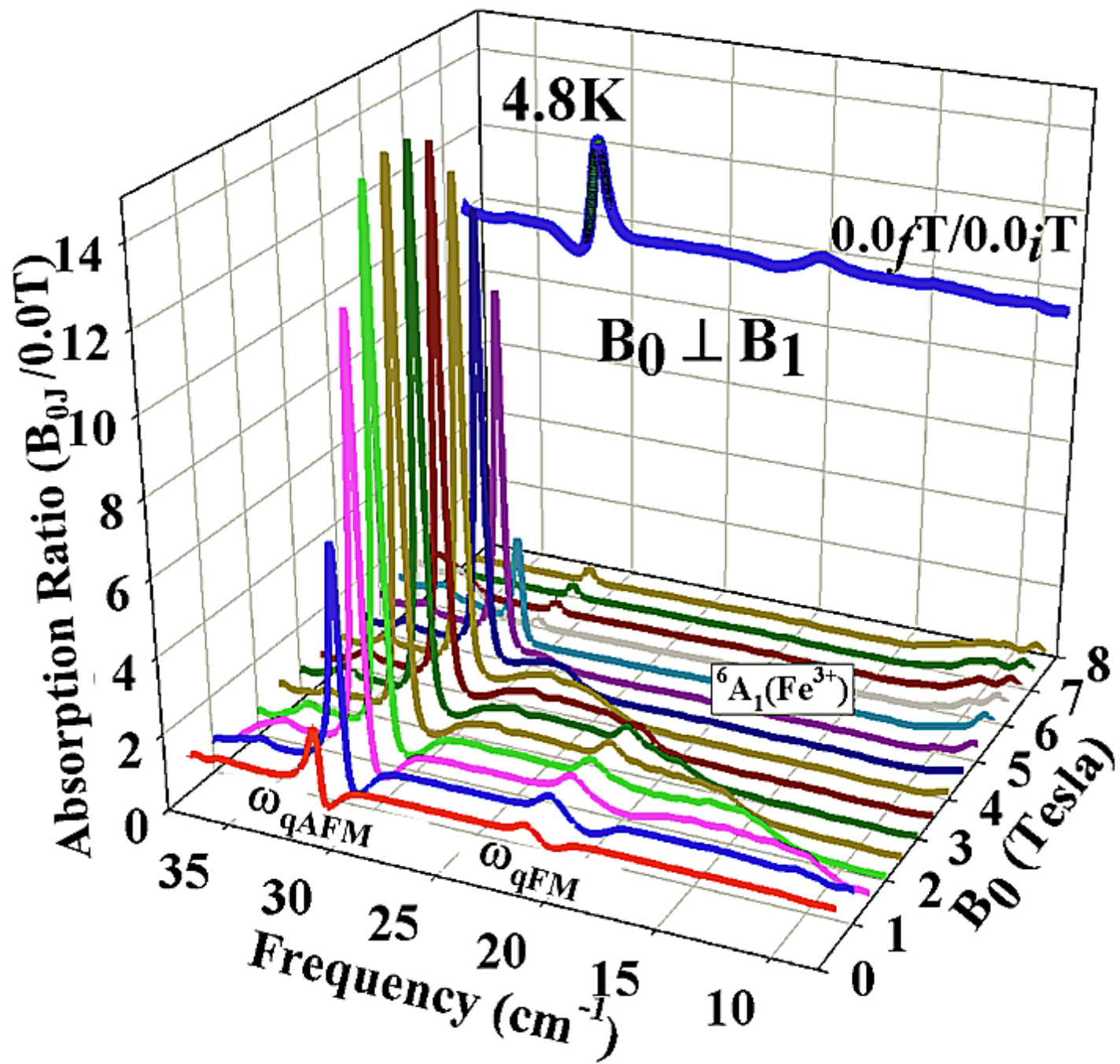

Fig. 9 MASSA et al

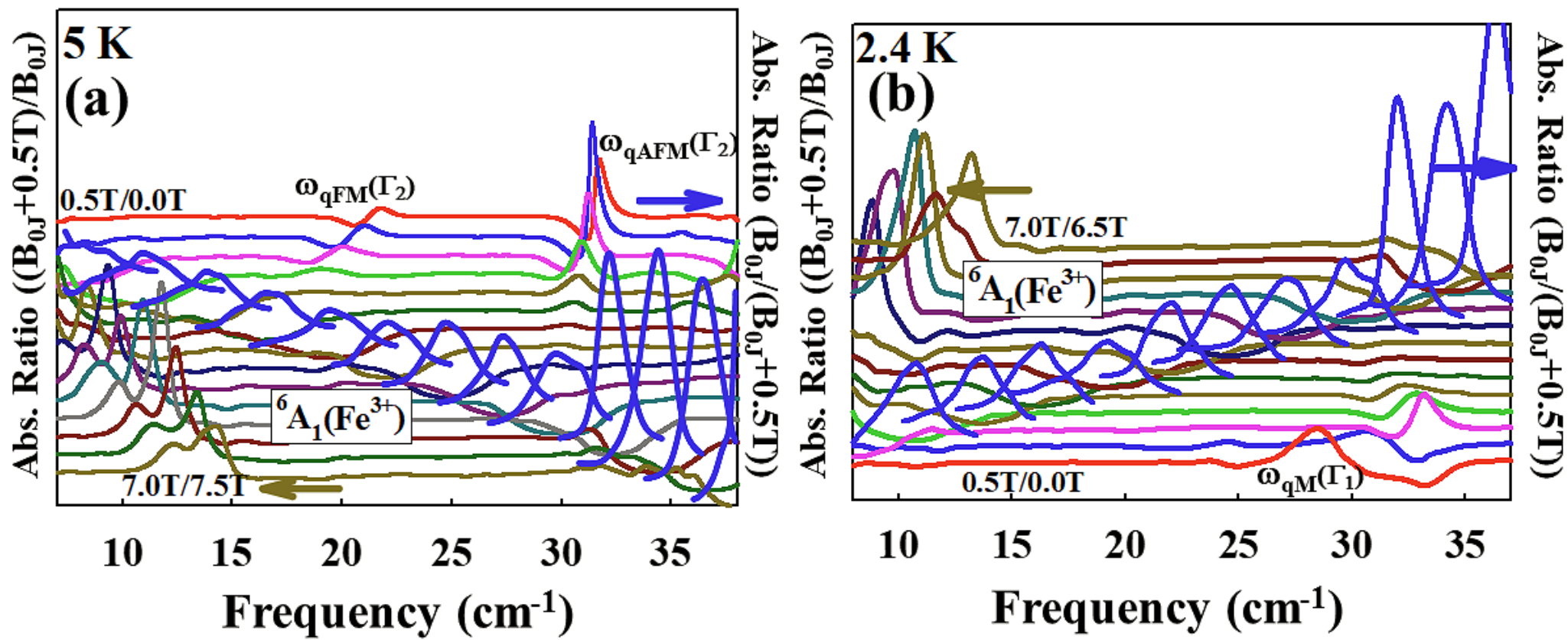

Fig. 10
MASSA et al

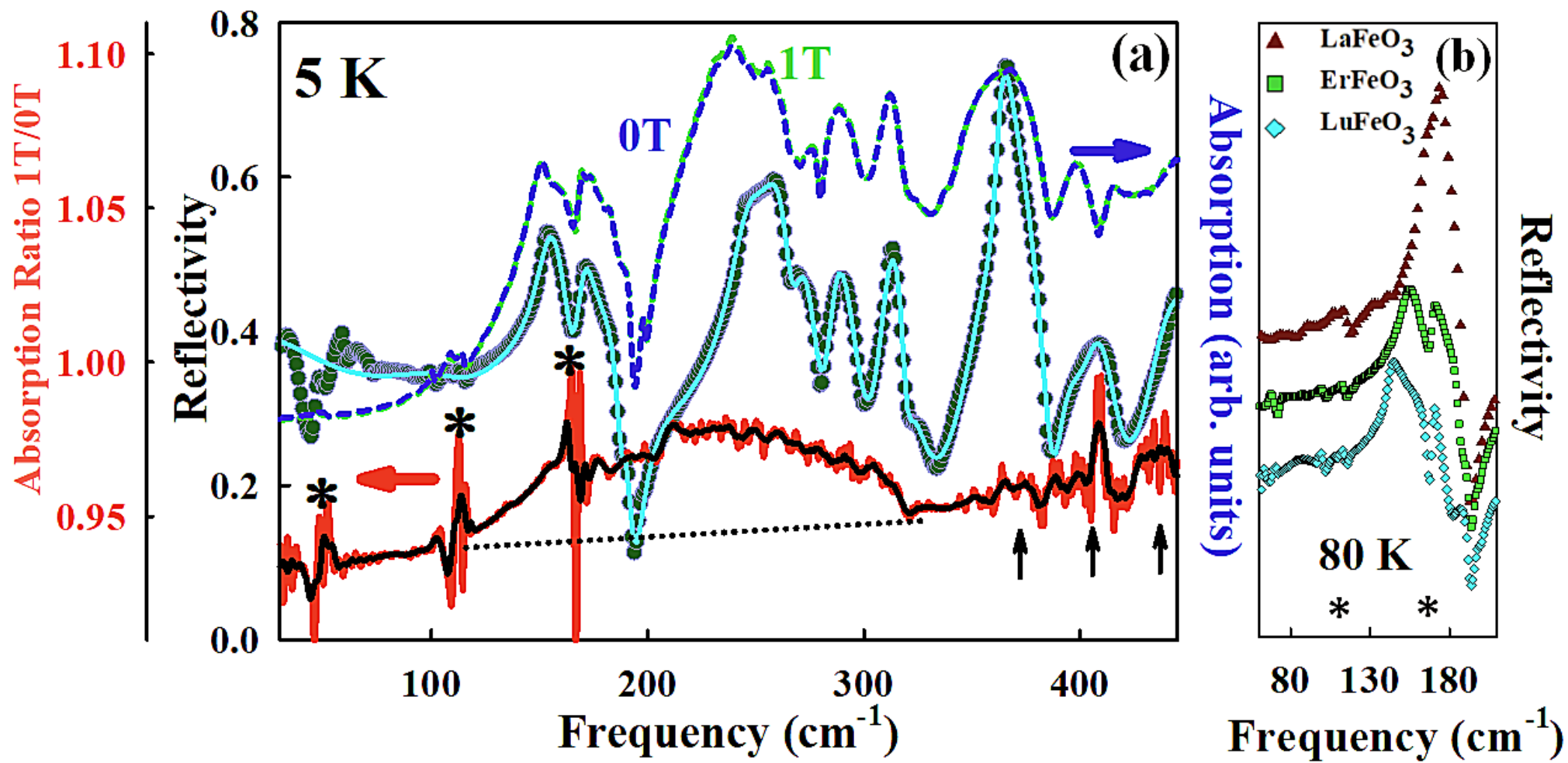

Fig. 11
MASSA et al

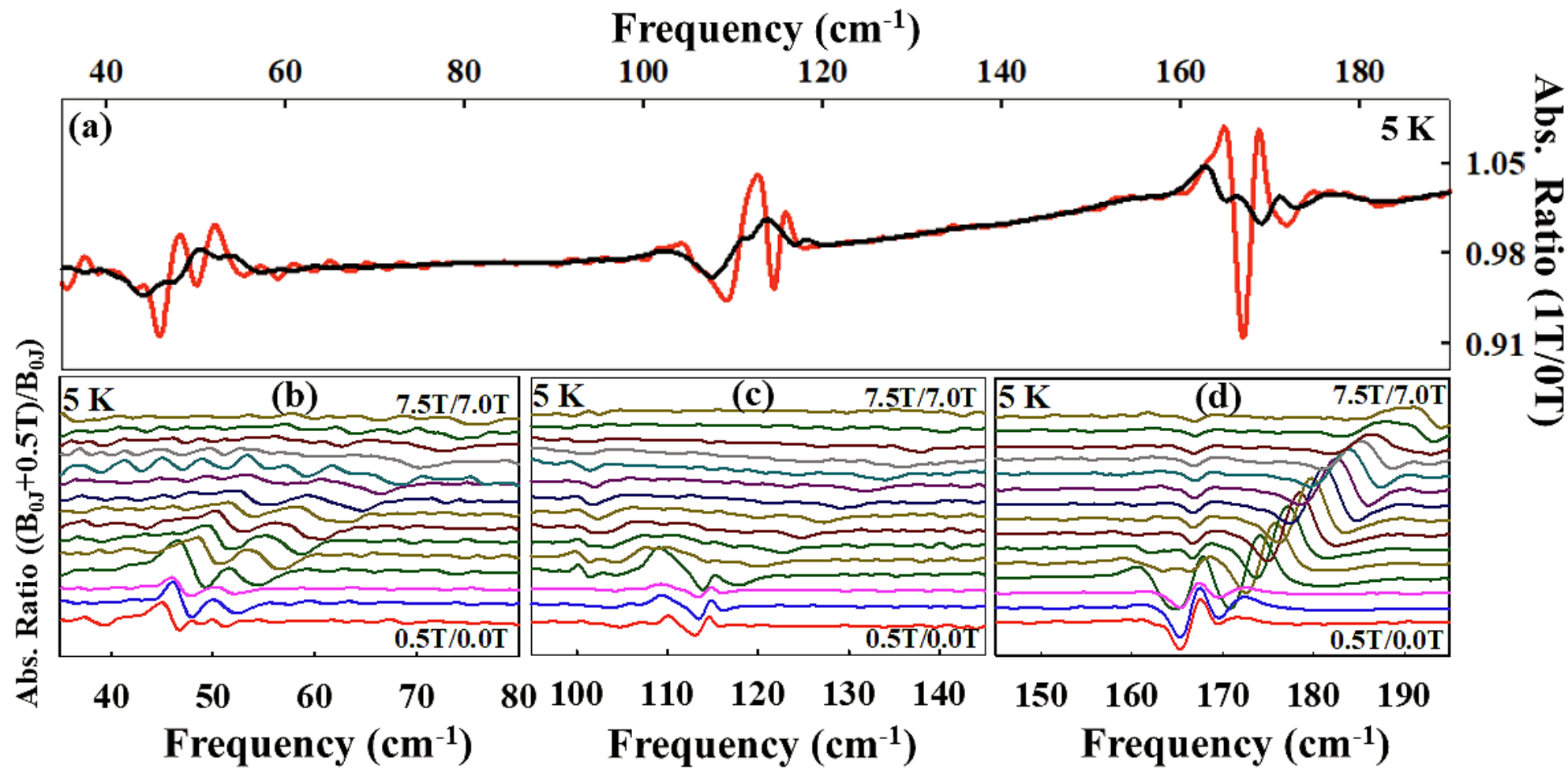

Fig. 12
MASSA et al

# SUPPLEMENTAL MATERIAL

# Low Temperature Terahertz Spectroscopy of LaFeO$_3$, PrFeO$_3$, ErFeO$_3$, and LuFeO$_3$: Quasimagnon resonances and ground multiplet transitions

Néstor E. Massa,*,[1] Leire del Campo,[2] Vinh Ta Phuoc,[3] Paula Kayser,[4‡] and José Antonio Alonso[5]

[1] Centro CEQUINOR, Consejo Nacional de Investigaciones Científicas y Técnicas, Universidad Nacional de La Plata, Bv. 120 1465, B1904 La Plata, Argentina.

[2] Centre National de la Recherche Scientifique, CEMHTI UPR3079, Université Orléans, F-45071 Orléans, France.

[3] Groupement de Recherche Matériaux Microélectronique Acoustique Nanotechnologies-UMR7347 CNRS, Université de Tours, INSA CVL, Parc Grandmont, F-37200, TOURS, France.

[4] Centre for Science at Extreme Conditions and School of Chemistry, University of Edinburgh, Kings Buildings, Mayfield Road, EH9 3FD Edinburgh, United Kingdom.

[5] Instituto de Ciencia de Materiales de Madrid, CSIC, Cantoblanco, E-28049 Madrid, Spain.

‡ Present address: Departamento de Química Inorgánica, Facultad de Ciencias Químicas, Universidad Complutense de Madrid, 28040 Madrid, Spain.

Corresponding author:
Néstor E. Massa *e-mail: neemmassa@gmail.com

# SAMPLE PREPARATION, MAGNETIC, and STRUCTURAL CHARACTERIZATION

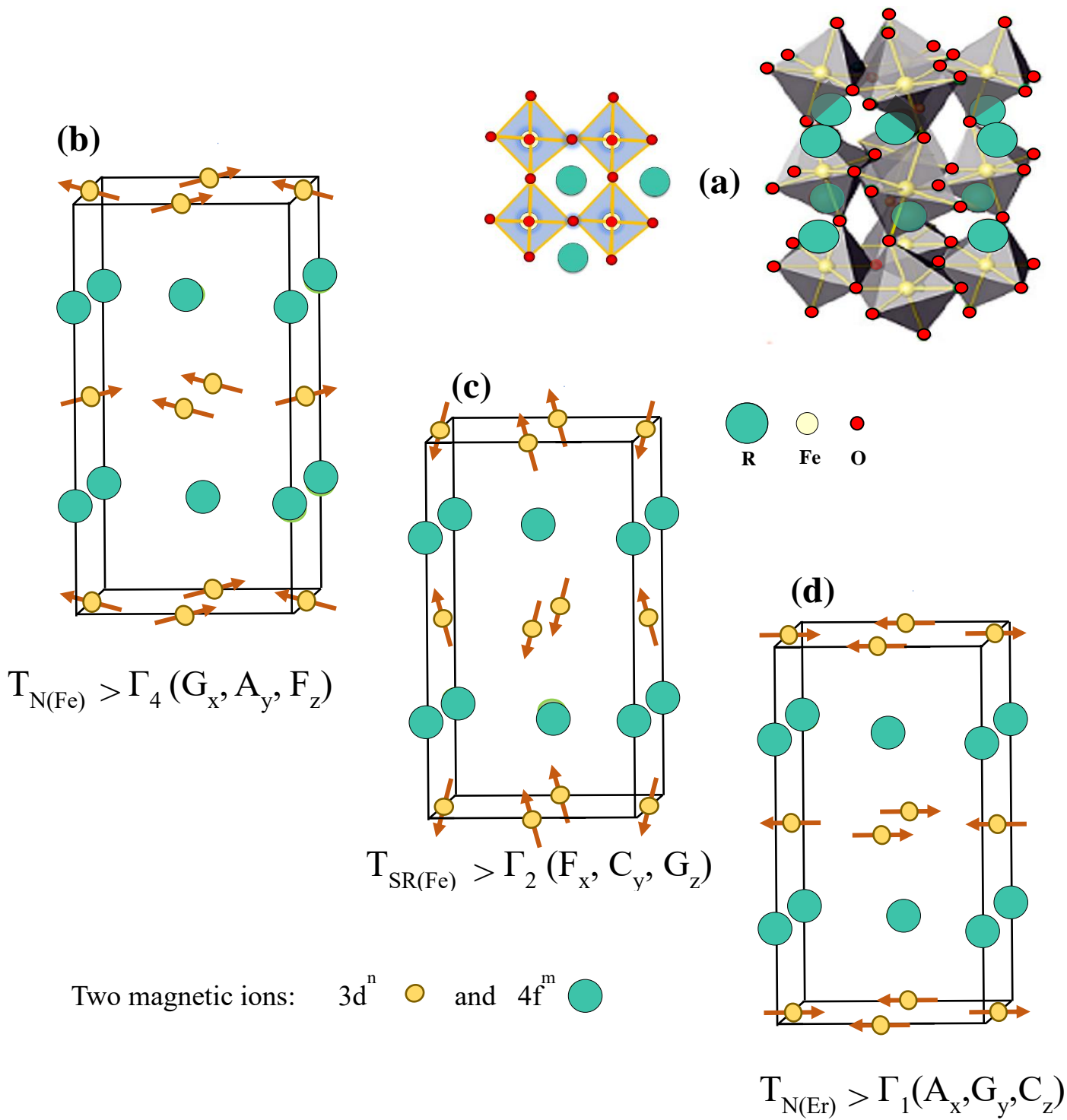

$T_{N(Fe)} > \Gamma_4 (G_x, A_y, F_z)$

$T_{SR(Fe)} > \Gamma_2 (F_x, C_y, G_z)$

$T_{N(Er)} > \Gamma_1 (A_x, G_y, C_z)$

Two magnetic ions: $3d^n$ and $4f^m$

RFeO$_3$ (R= La, Pr, Er, Lu) polycrystalline samples were prepared by standard ceramic synthesis procedures. Stoichiometric amounts of analytical grade Fe$_2$O$_3$ and R$_2$O$_3$ powder oxides were thoroughly ground and heated in air at 1000ºC for 12 h and 1300ºC for 12 h in alumina crucibles. Then, pellets of ~1cm diameter, less than 2mm thick, were prepared by uniaxial pressing the raw powders and sintering the disks at 1300ºC for 2 h. The purity of the samples for all four RFeO$_3$ (R= La, Pr, Er, Lu) was checked by X-ray powder diffraction (XRD) collected at room temperature with Cu-Kα radiation. Shown in Fig. S2, all data were analyzed using the Rietveld method with refinements carried out with the program FULLPROF.

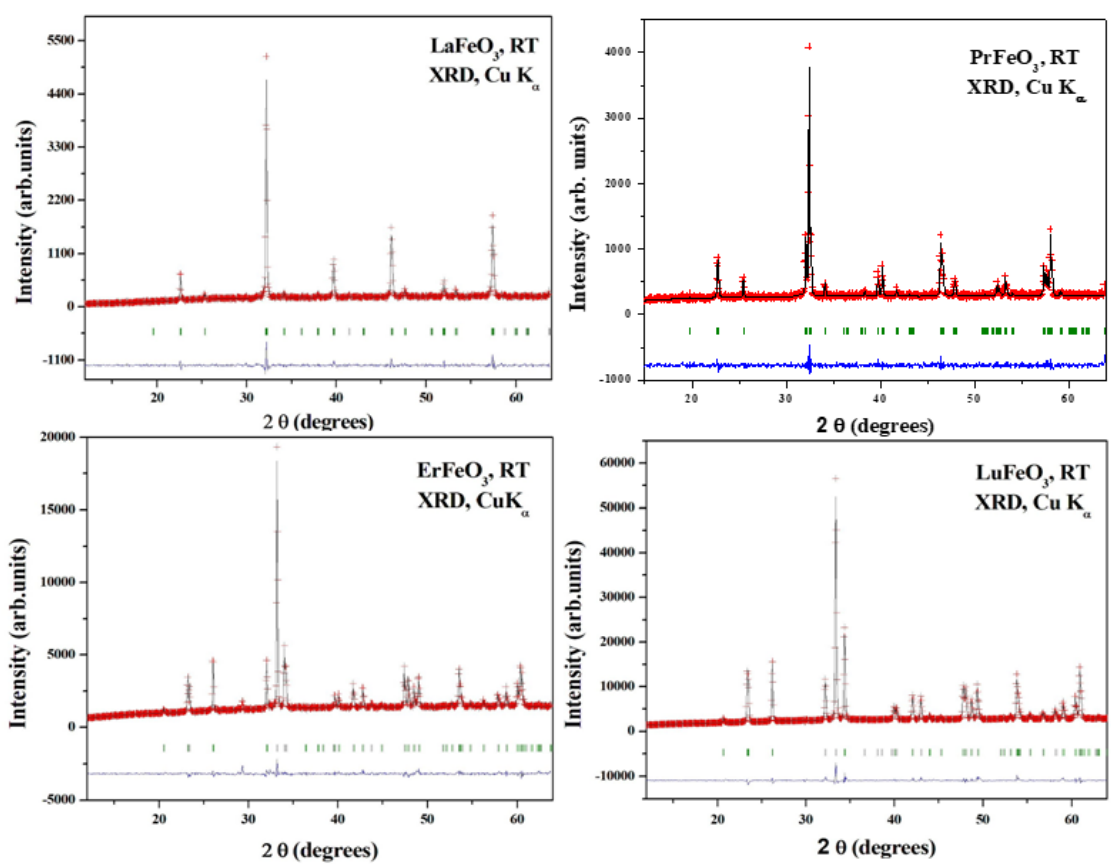

**Fig. S2** Fitted X-ray (CuKα) diffraction patterns of LaFeO$_3$, PrFeO$_3$, ErFeO$_3$, and LuFeO$_3$ in the room temperature crystalline space group P$_{bmn}$ (D$_{2h}^{16}$) with four molecule per unit cell (Z=4). Vertical bars are reflection positions of the orthorhombic unit cell used in the Rietveld refinement. The bottom curve is the difference between the experimental and calculated patterns.

# EXPERIMENTAL METHODS AND DATA ANALYSIS

## a) *THz measurements*

Low temperature-low frequency absorption measurements from 3 cm$^{-1}$ to 50 cm$^{-1}$ with 0.5 resolution have been performed in the THz beamline in the low-alpha multi bunch hybrid mode of the BESSY II storage ring at the Helmholtz-Zentrum Berlin (HZB) [1] In the synchrotron low-alpha mode electrons are compressed within shorter bunches of only ~2 ps duration allowing far-infrared wave trains up to mW average power to overlap coherently in the THz range below 50 cm$^{-1}$. [2,3] All measurements have been done in the Voigt configuration.

To avoid possible artifacts introduced in our spectra by the beam operation our measurements are presented as spectrum ratios, as in fig. 3, in which we normalized lower temperature spectra by a higher temperature spectrum aiming to mostly compensate extrinsic anomalies. For the same reason we use absorption incremental sequential ratios on field dependent measurements (Abs. Ratio (($B_{0J}$+0.5T)/$B_{0J}$), $B_{0j}$ is the applied field at the j$^{th}$ incremental step, 0.5 is the field step increment in Teslas) since by doing this we were able to verify the absence of run glitches (e.g., fig.7) besides checking the compound magnetic response to the field step increments. This, in turn, allows to now calculate with confidence band features that are induced only by the applied field (e.g., fig. 4), here, using the ZFC spectrum as normalizing reference.

Polyethylene pellets embedded with ErFeO$_3$ powder were used in the far infrared measurements done with the beamline interferometer Bruker IFS125 HR.

We used a superconducting magnet (Oxford Spectromag 4000, here up 7.5 T) interfaced with the interferometer Bruker IFS125 HR for the measurements under magnetic fields. The temperature was measured with a calibrated Cernox Sensor from LakeShore Cryotronics mounted to the copper block that holds the sample in the Variable Temperature Insert (VTI) of the Spectromag 4000 Magnet. Measurements in the 30 cm$^{-1}$ to 800 cm$^{-1}$ range were also taken using the internal source of the Bruker IFS125 HR. [2]

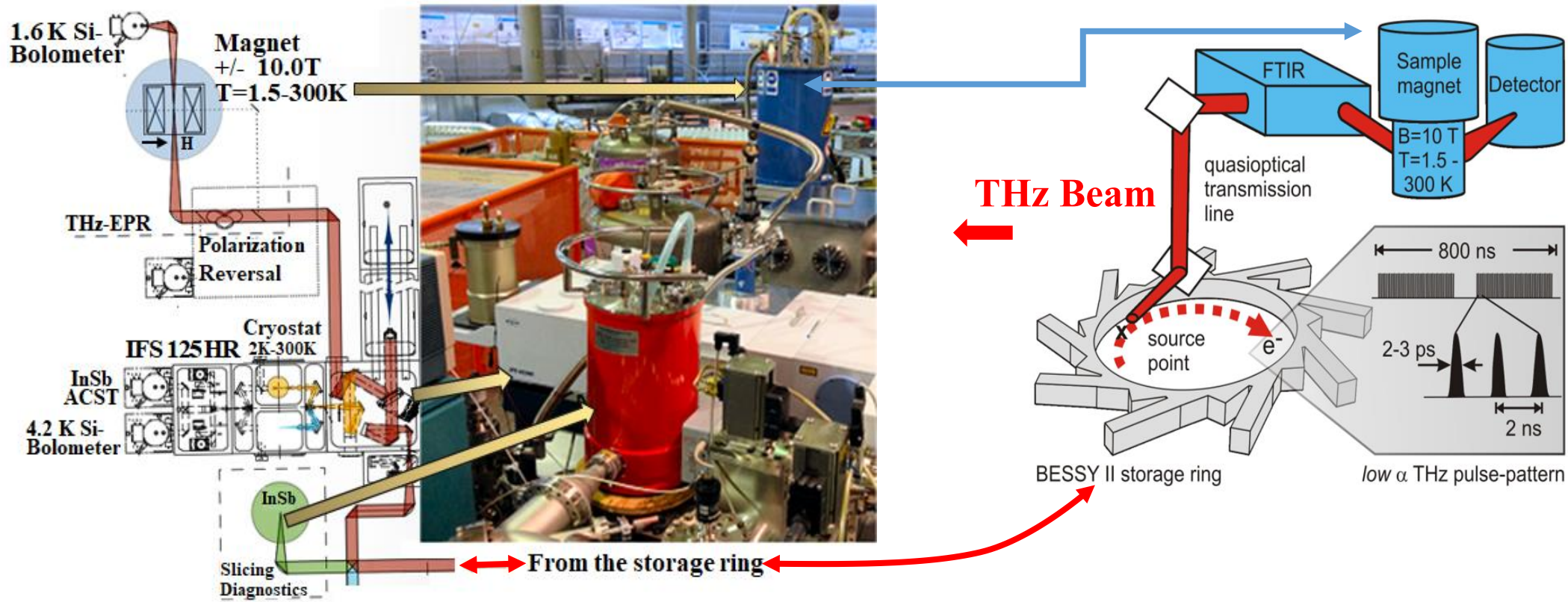

**Fig. S3**. Beam optical path and layout of the THz beamline at BESSY II [1-3]

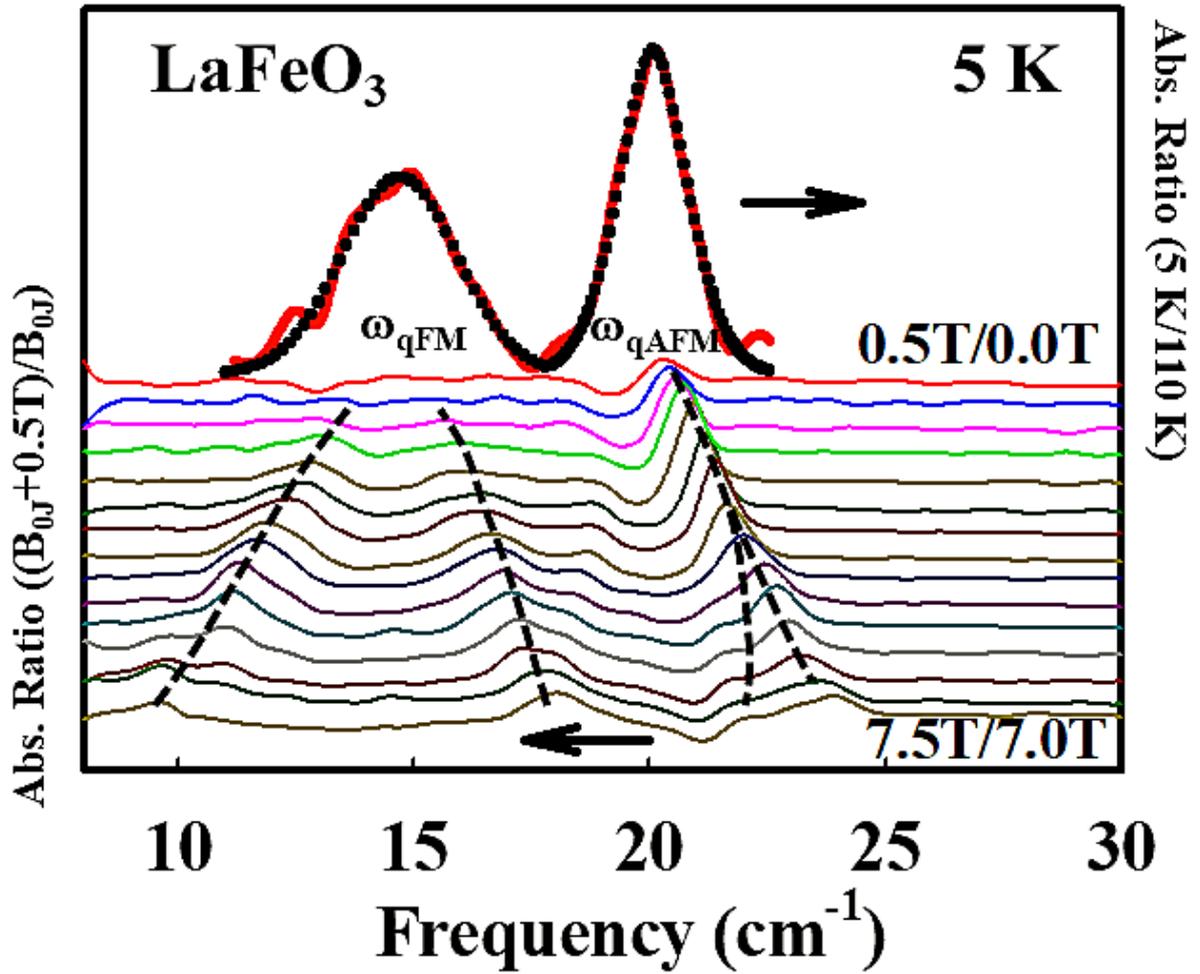

**Fig. S4.** Upper panel: LaFeO$_3$ 5K zone center quasiantiferromagnetic ($\omega_{qAFM}$) and quasiferromagnetic ($\omega_{qFM}$) spin wave resonances; full line: measurement, dots: Gaussian fit. Lower panel: sequential absorption ratios, (B$_{0j}$+0.5T)/B$_{0j}$, B$_{0j}$ is the applied field at the j$^{th}$ incremental step, 0.5 is the field step increment in Teslas. Dashed lines are guides for the eye.

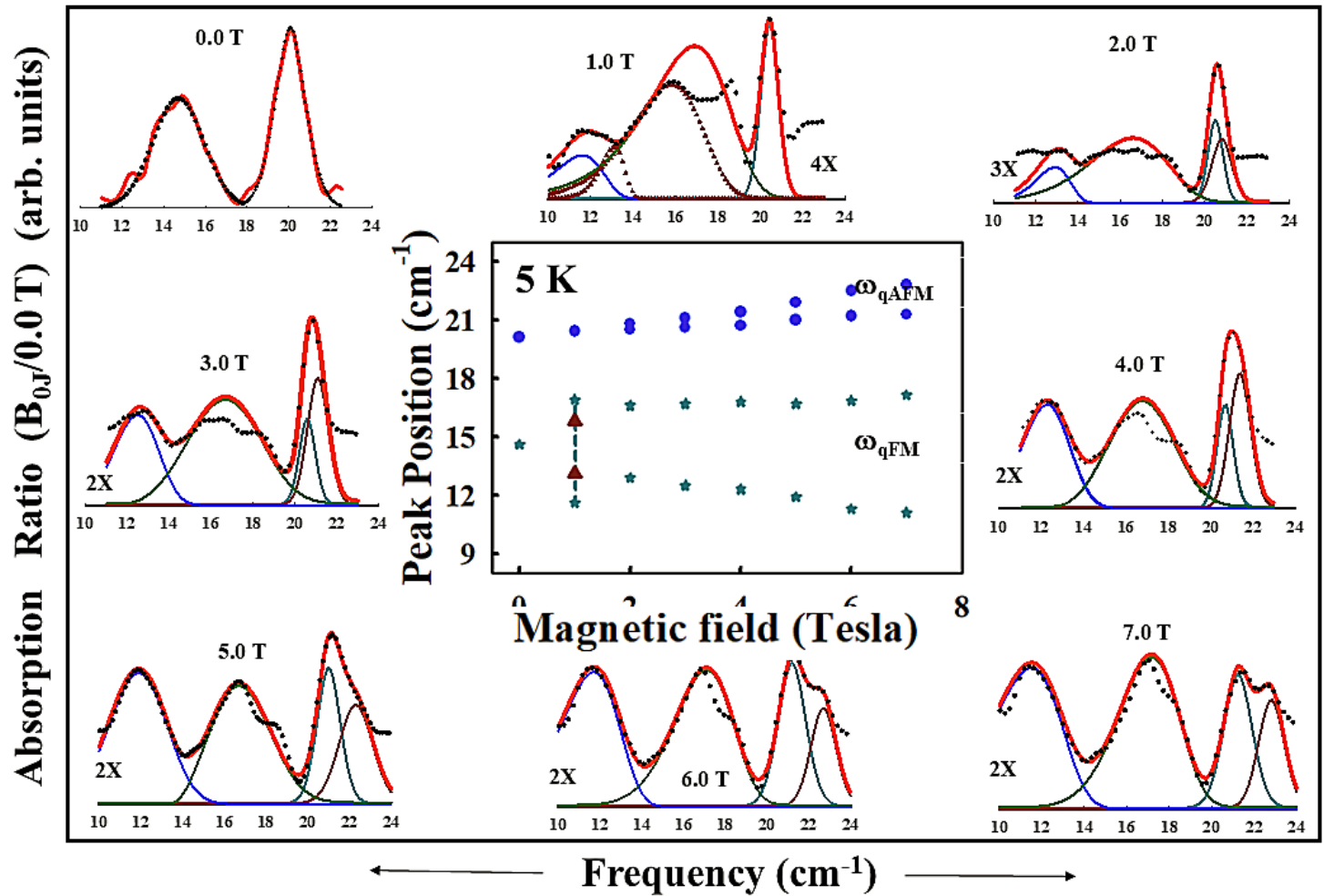

**Fig. S5.** LaFeO$_3$ quasiantiferromagnetic ($\omega_{qAFM}$) and quasiferromagnetic ($\omega_{qFM}$) spin wave resonances as function of the applied magnetic fields. Insets: zero field cooled 0.0 T absorption and absolute field induced absorption resonance ratios from 1.0T to 7.0T at 5 K, dots: measurement, full lines: Gaussian and Weibull fits. The bands delineated with upper triangles in the 1.0T inset are the profile main constituents centered at 15.8 cm$^{-1}$ and 13.1 cm$^{-1}$. Center inset: peak positions of the shown absorptions.

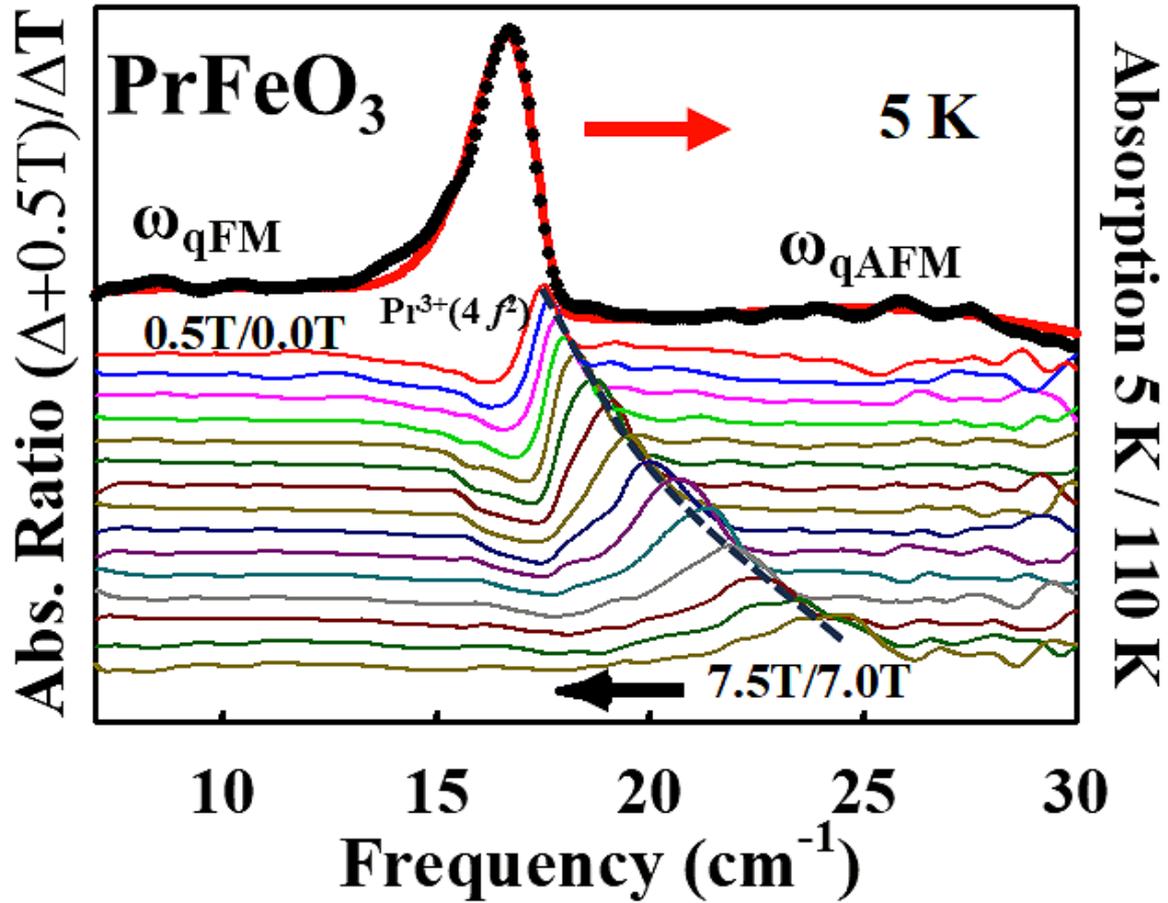

**Fig. S6** Upper panel: $Pr^{3+}(4f^2)$ $^3H_4$ ground state singlet and poorly defined bands assigned to $\omega_{qFM}$ and $\omega_{qAFM}$ magnon profiles at 5 K in $PrFeO_3$ (see text); full line: measurement, dots: Gaussian and Weibull fits. Lower panel: sequential absorption ratios, $(B_{0j}+0.5T)/B_{0j}$, $B_{0j}$ is the applied field at the $j^{th}$ incremental step, 0.5 is the field step increment in Teslas. Dashed line is a guide for the eye.

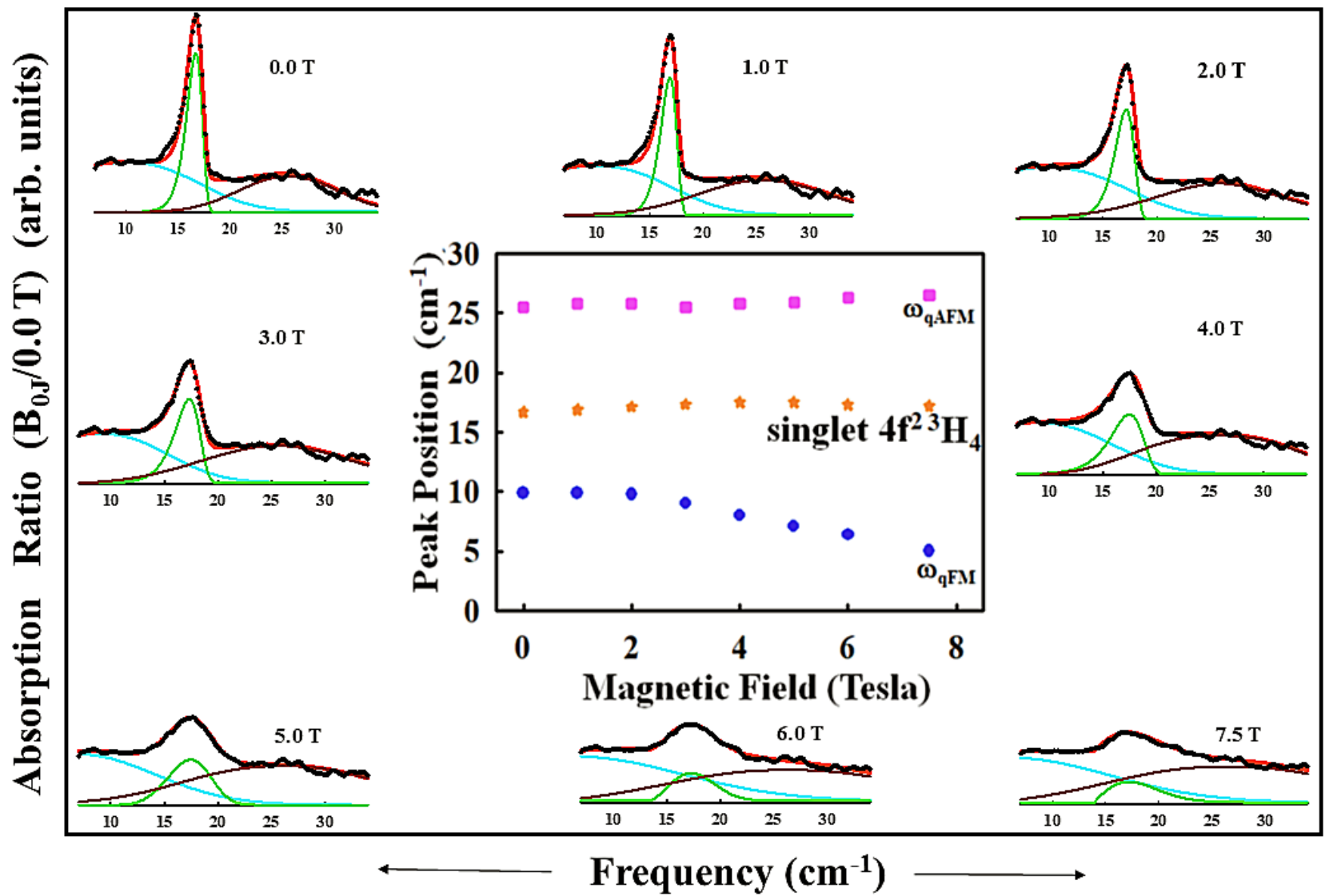

**Fig. S7** Singlet lowest crystal field level of the Pr$^{3+}$ ($^3H_4$) manifold at selected magnetic fields and Gaussian deconvolutions of the broader and weaker side bands assigned to the quasiferromagnetic ($\omega_{qFM}$) and quasiantiferromagnetic ($\omega_{qAFM}$) resonances at 5 K. Center inset: peak positions of the shown absorptions

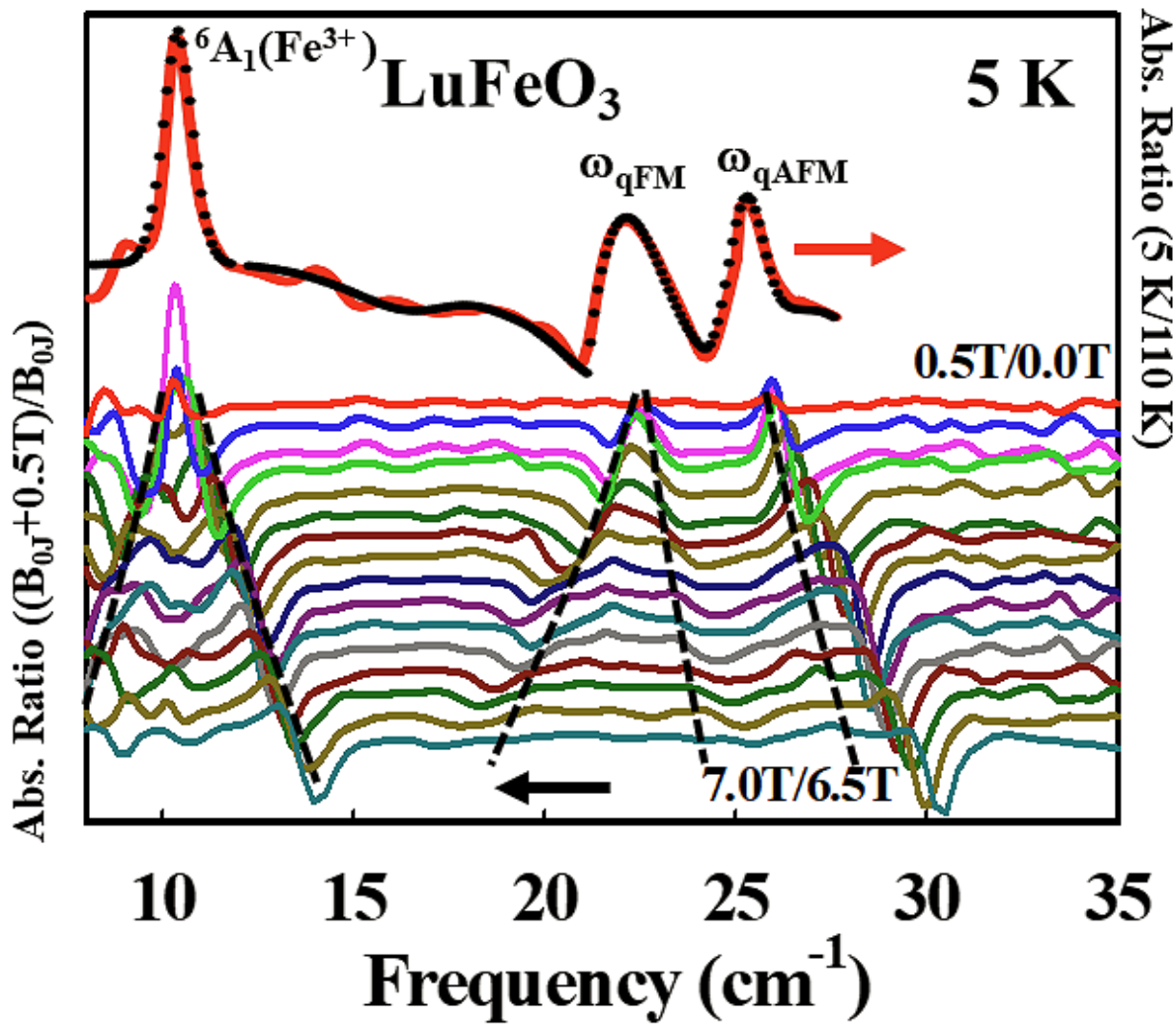

**Fig. S8**. Upper panel: Zone center $Fe^{3+}$ ($^6A_1$) crystal field transition and quasiantiferromagnetic ($\omega_{qFM}$) and quasiferromagnetic ($\omega_{qAFM}$) magnon modes of $LuFeO_3$ at 5 K; full line: measurements, dots: Gaussian fits. Lower panel: sequential absorption ratios, $(B_{0j}+0.5T)/B_{0j}$, $B_{0j}$ is the applied field at the $j^{th}$ incremental step, 0.5 is the field step increment in Teslas. Dashed lines are guides for the eye.

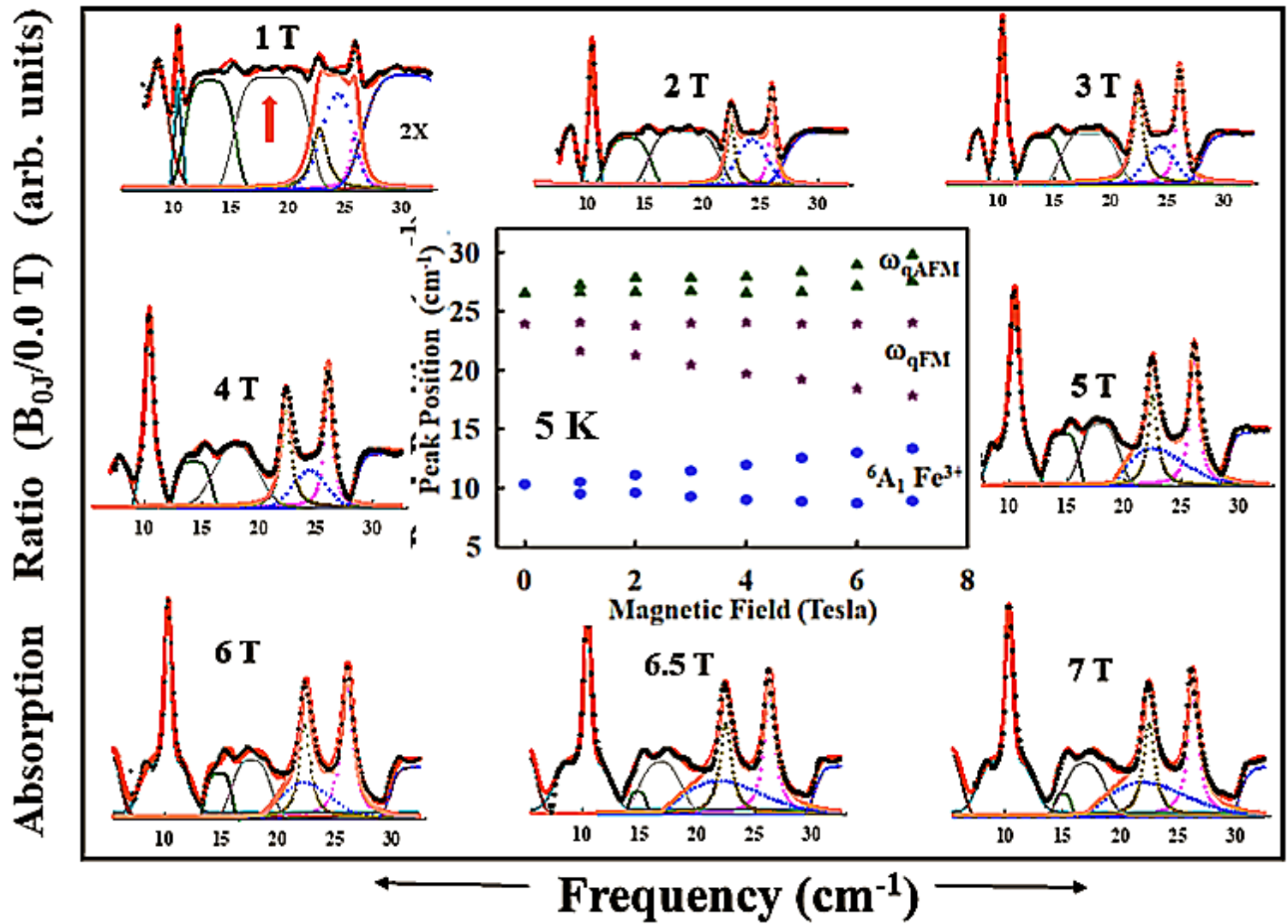

**Fig. S9**. $Fe^{3+}$ ($^6A_1$) crystal field transition and quasiferromagnetic ($\omega_{qFM}$) and quasiantiferromagnetic ($\omega_{qAFM}$) resonances of $LuFeO_3$ as function of the applied magnetic field. Insets: absolute field induced absorption ratios for $LuFeO_3$ from 1.0T to 7.0T at 5 K, dots: measurement; full lines: Gaussian and Weibull fits. The arrow at 1T signals secondary oscillations attributed to individual magnetic site contributions out of the two-site approximation. Center inset: peak positions of the shown absorptions.

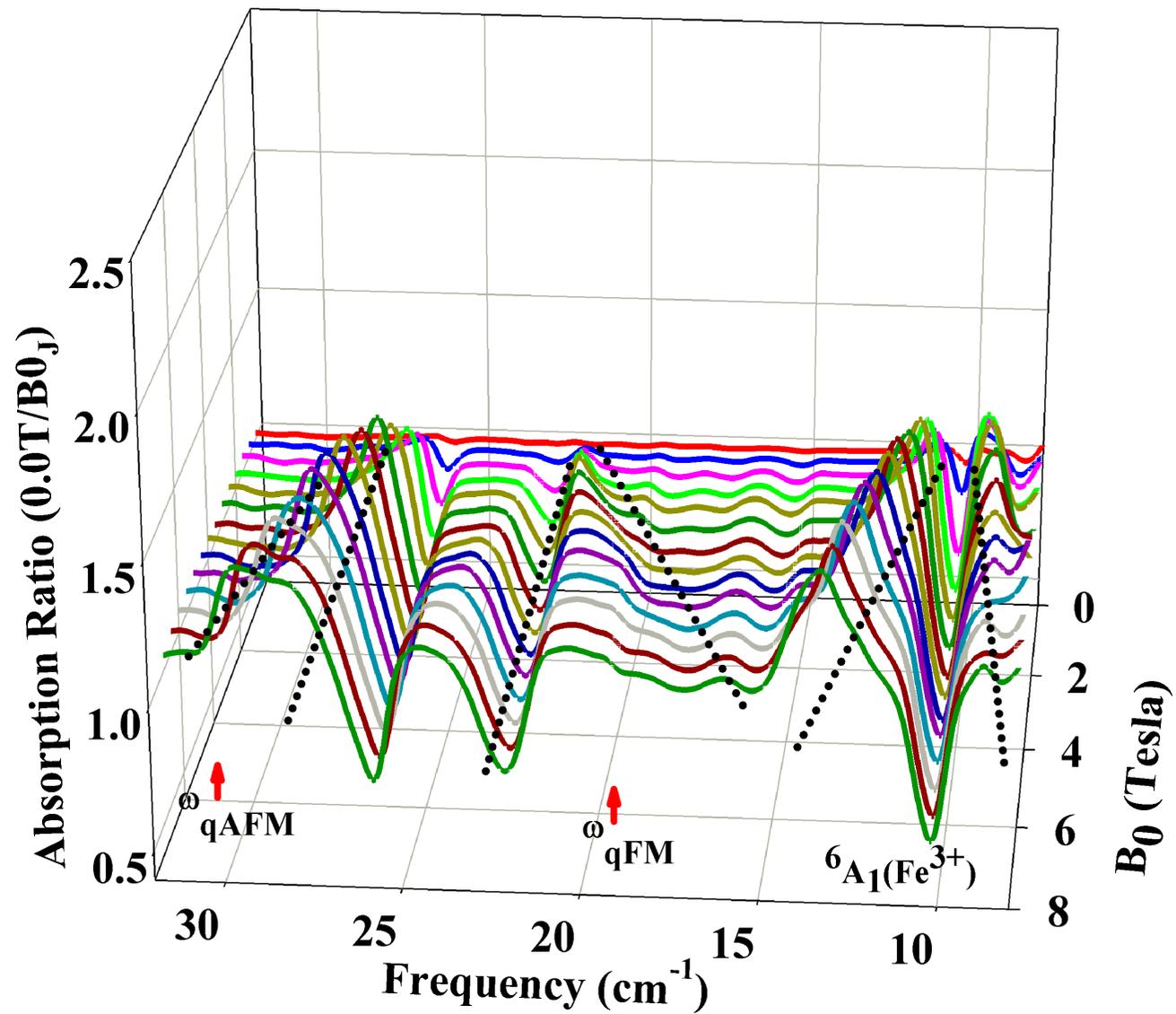

**Fig. S10** Reciprocal view highlighting the field induced band split in the absorption ratios ($B_{0J}/0.0T$) of LuFeO$_3$ at 5 K.

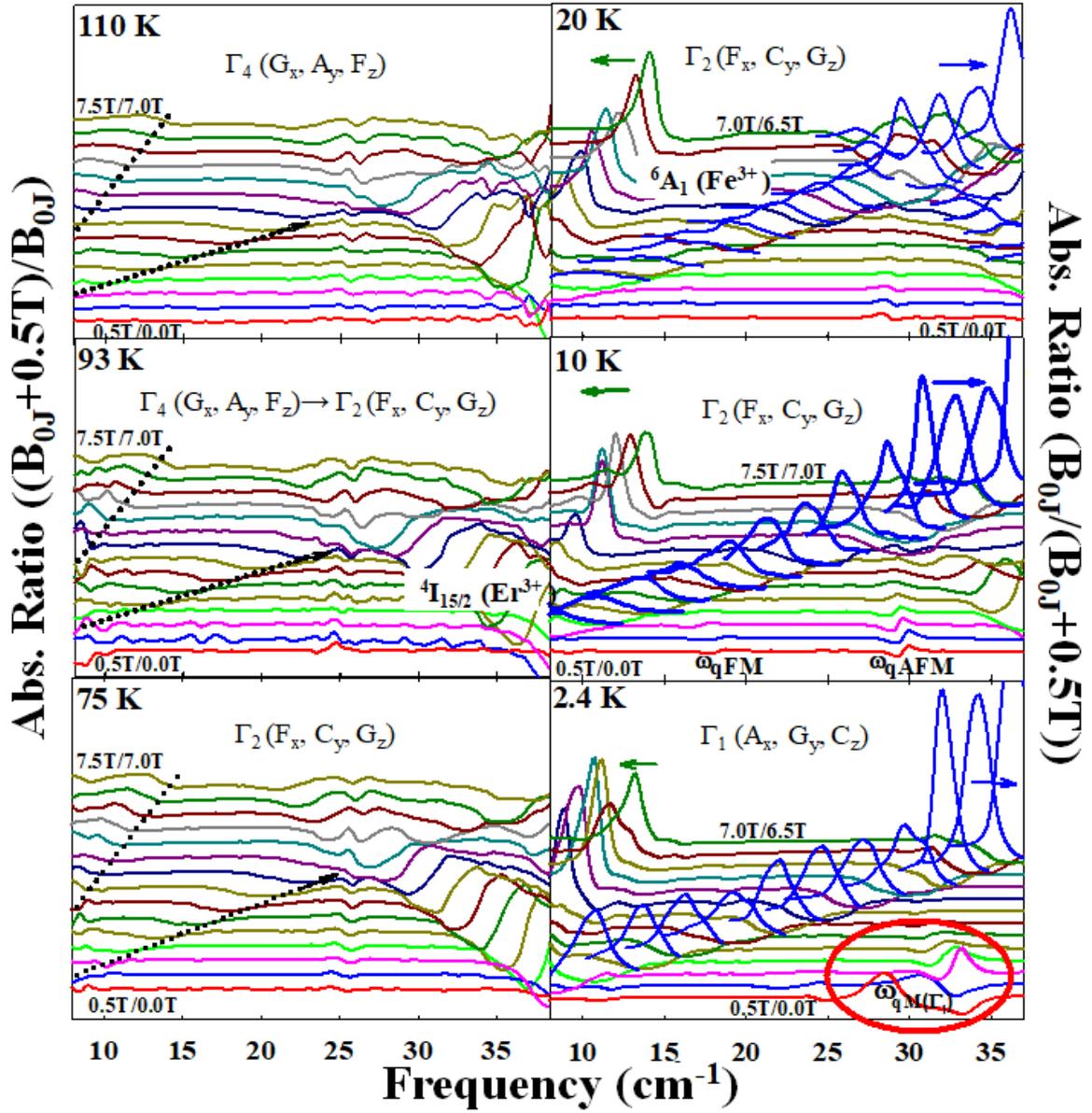

**Fig. S11.** Sequential absorption ratio spectra, $(B_{0j}+0.5T)/B_{0j}$, $B_{0j}$ is the applied field at the $j^{th}$ incremental step, of ErFeO$_3$ crystal field Zeeman splits of Fe$^{3+}$ ($^6A_1$) and Er$^{3+}$ ($^4I_{15/2}$) multiplets and quasimagnon modes from 110 K to 2.4 K in the $\Gamma_4$ (G$_x$, A$_y$, F$_z$), $\Gamma_2$ (F$_x$, C$_y$, G$_z$), $\Gamma_1$ (A$_x$, G$_y$, C$_z$) zero field magnetic phases. As it is discussed in the main text, the two resonances merge in the $\Gamma_1$ phase into one weaker broader single mode, shown encircled at 2.4 K, with a strong half-width at half-maximum field dependence.

## a) *Far infrared measurements*

Far infrared near normal reflectivity spectra using conventional near normal incidence geometry were taken at the CEMHTI (Conditions Extrêmes et Matériaux: Haute Température et Irradiation- UPR3079 CNRS-Orléans) facilities on heating from 4 K to 300 K at 1 cm$^{-1}$ resolution with a Bruker 113V interferometer and at GREMAN (Groupement de Recherche Matériaux Microélectronique Acoustique Nanotechnologies- UMR 7347, Université François Rabelais Tours. Samples were mounted on the cold finger of a He- closed cycle refrigerator and a home-made He cryostat adapted to the near normal reflectivity attachment vacuum chamber of the interferometer. A liquid He cooled bolometer and a deuterated triglycine sulfate pyroelectric bolometer (DTGS) were employed to completely cover the spectral range of interest. A plain gold mirror and an in-situ evaporated gold film were used for 100% reference reflectivity.

We analyzed the reflectivity spectra using the standard procedures for multioscillator dielectric simulation given by,

$$\varepsilon(\omega) = \varepsilon_1(\omega) - i\varepsilon_2(\omega) = \varepsilon_\infty \prod_j \frac{(\omega_{jLO}^2 - \omega^2 + i\gamma_{jLO}\omega)}{(\omega_{jTO}^2 - \omega^2 + i\gamma_{jTO}\omega)} \quad (1)$$

where $\varepsilon_1(\omega)$ is the real and the $\varepsilon_2(\omega)$ imaginary part of the dielectric function (complex permittivity, $\varepsilon^*(\omega)$), $\varepsilon_\infty$ is the high frequency dielectric constant taking into account electronic contributions; and where $\omega_{jTO}$ and $\omega_{jLO}$, are the transverse and longitudinal optical mode frequencies and with $\gamma_{jTO}$ and $\gamma_{jLO}$ their respective damping.[4]

We then fitted at every temperature the experimental near normal reflectivity by computing each with

$$R(\omega) = \left| \frac{\sqrt{\varepsilon^*(\omega)} - 1}{\sqrt{\varepsilon^*(\omega)} + 1} \right|^2 \quad (2)$$

This allowed to extract the experimental phonon frequencies that were then matched to feature bands of the corresponding absorption spectrum.[5] With few exceptions our measurements agree with the 25 allowed infrared active zone center phonons [6, 7, 8] predicted for a compound with 4 molecules per unit cell (Z=4) sustaining the Pbnm space group. The irreducible representation is given by

$$\Gamma_{IR} = 7B_{1u} + 9B_{2u} + 9B_{3u} \qquad (3)$$

A broad feature appearing in the reflectivity of ErFeO$_3$ and LuFeO$_3$ at our lowest measuring limit assigned to a boson peak has been Gaussian fitted according

$$G = A * e^{-\sqrt{(x-Xc)}}/w \qquad (4)$$

where A is the amplitude, Xc is its peak position, and w its width.[9]

## LaFeO$_3$ - Reflectivity and Multioscillator Fitting Parameters

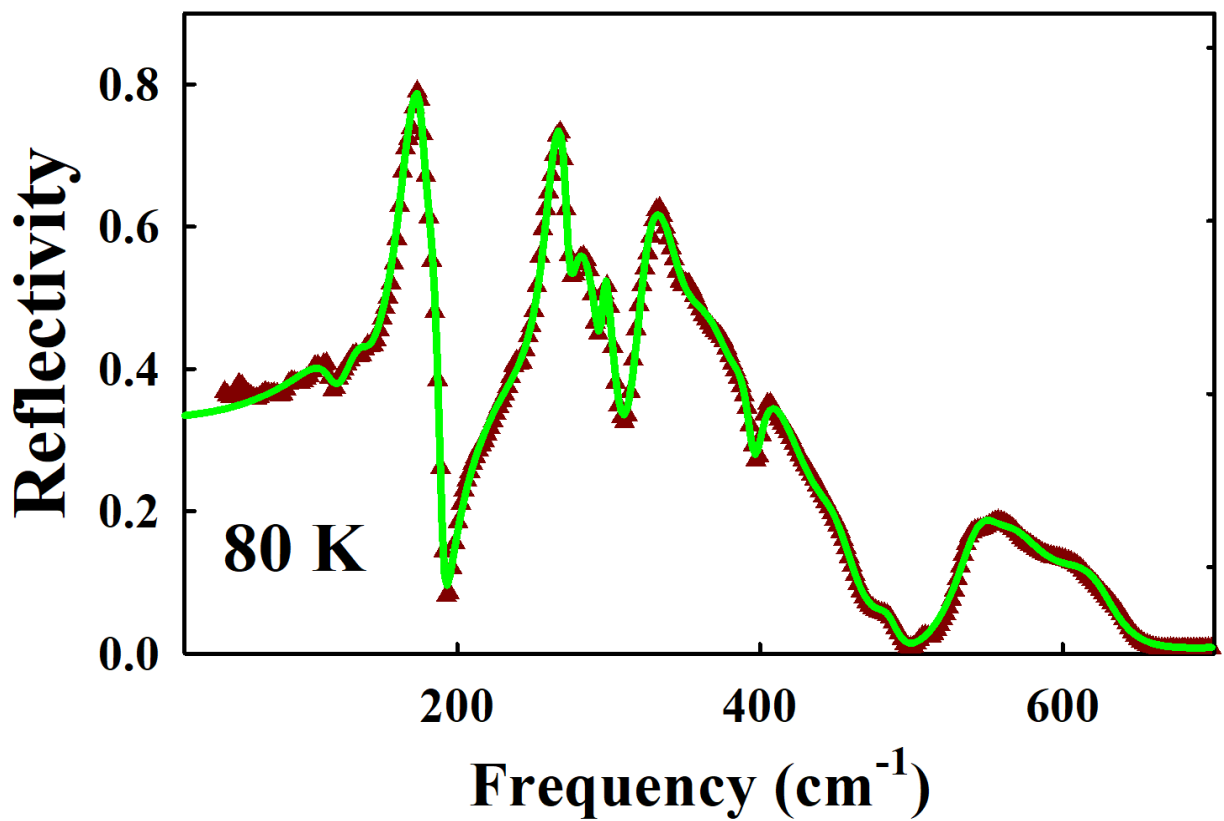

**Fig. S12.** LaFeO$_3$ near normal reflectivity at 80 K; experimental: upper triangles (showing 1 out of 10 points), full line: fit.

# TABLE SI

Dielectric simulation fitting parameters for LaFeO$_3$ reflectivity at 80 K.
.

| T (K) | $\varepsilon_\infty$ | $\omega_{TO}$ (cm$^{-1}$) | $\Gamma_{TO}$ (cm$^{-1}$) | $\omega_{LO}$ (cm$^{-1}$) | $\Gamma_{LO}$ (cm$^{-1}$) |
|---|---|---|---|---|---|
| 80 | 2.65 | 117.0 | 27.4 | 119.7 | 21.0 |
| | | 137.1 | 21.0 | 140.3 | 20.7 |
| | | 164.8 | 13.1 | 117.8 | 6.3 |
| | | 177.1 | 13.2 | 191.1 | 10.1 |
| | | 231.8 | 74.9 | 254.8 | 61.5 |
| | | 261.0 | 12.7 | 273.0 | 7.9 |
| | | 275.1 | 13.6 | 285.9 | 8.4 |
| | | 296.5 | 5.4 | 306.0 | 23.2 |
| | | 321.2 | 17.2 | 344.9 | 36.3 |
| | | 350.7 | 46.0 | 381.9 | 29.2 |
| | | 384.5 | 27.4 | 395.1 | 12.3 |
| | | 397.9 | 14.3 | 437.2 | 69.4 |
| | | 445.8 | 40.7 | 467.5 | 40.5 |
| | | 484.5 | 23.0 | 488.4 | 19.4 |
| | | 540.0 | 29.6 | 558.2 | 35.7 |
| | | 562.0 | 33.5 | 573.1 | 83.7 |
| | | 615.8 | 57.3 | 630.8 | 42.8 |

# ErFeO₃ Reflectivity and Dielectric Simulation Fitting Parameters

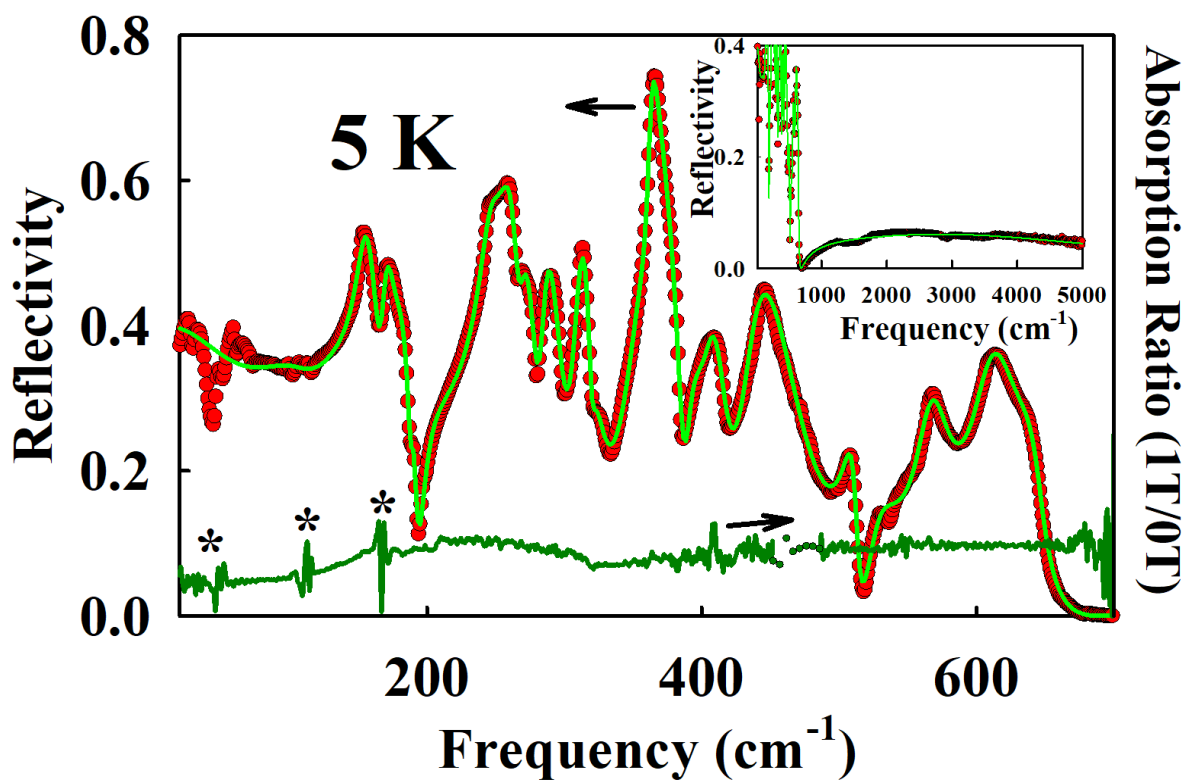

**Fig. S13.** Orthorhombic ErFeO$_3$ near normal reflectivity at 5 K, experimental: dots (showing 1 out of 10 points), full line: fit. Inset: full measured reflectivity range. The absorption ratio between 0 T and 1T spectra at 5 K is shown to emphasize the absence of a magnetic induced response at ~600 cm$^{-1}$ breathing mode frequencies. Asterisks denote the Er$^{3+}$ ($^4I_{15/2}$) crystal field energies.

# Table SII

Dielectric simulation fitting parameters for ErFeO$_3$ reflectivity at 5 K. Bottom cells show the parameters used for the lower frequency Gaussian fit. Phonon frequencies appearing in the fits likely related to Er$^{3+}$ ($^4$I$_{15/2}$) crystals field energies are indicated in bold italics with an asterisk.

| (K) | $\varepsilon_\infty$ | $\omega_{TO}$ (cm$^{-1}$) | $\Gamma_{TO}$ (cm$^{-1}$) | $\omega_{LO}$ (cm$^{-1}$) | $\Gamma_{LO}$ (cm$^{-1}$) |
|---|---|---|---|---|---|
| 5 | 1.24 | ***110.5*** | ***50.2*** | ***114.6*** | ***45.8*** |
| | | 154.0 | 11.9 | 156.4 | 34.7 |
| | | 156.0 | 73.7 | 166.0 | 10.7 |
| | | ***169.2*** | ***7.6*** | ***175.7*** | ***19.6*** |
| | | 179.2 | 19.8 | 187.8 | 4.5 |
| | | 188.9 | 4.7 | 193.0 | 6.7 |
| | | 197.6 | 26.5 | 207.4 | 76.1 |
| | | 244.5 | 14.3 | 250.9 | 23.5 |
| | | 257.0 | 19.0 | 265.9 | 9.8 |
| | | 269.1 | 12.0 | 280.0 | 11.8 |
| | | 285.5 | 9.7 | 299.9 | 20.0 |
| | | 309.8 | 10.6 | 319.3 | 6.0 |
| | | 320.3 | 8.4 | 326.3 | 27.9 |
| | | 353.5 | 21.7 | 362.2 | 12.4 |
| | | 361.8 | 6.2 | 386.1 | 10.7 |
| | | 390.2 | 11.8 | 399.6 | 38.9 |
| | | 406.4 | 18.4 | 415.4 | 23.6 |
| | | 434.2 | 24.4 | 474.2 | 71.5 |
| | | 509.2 | 18.8 | 513.2 | 7.6 |
| | | 524.7 | 21.1 | 527,6 | 41.3 |
| | | 562.3 | 24.6 | 573.1 | 34.4 |
| | | 603.4 | 37.4 | 620.1 | 41.4 |
| | | 642.3 | 80.3 | 644.7 | 14.4 |
| | | 3731.9 | 4987.3 | 5438 | 8486.3 |
| | | A | Xc | w | |
| | | 470 | 11.1 | 30.5 | |

**ErFeO$_3$ -Reflectivity and Multioscillator Fitting Parameters**

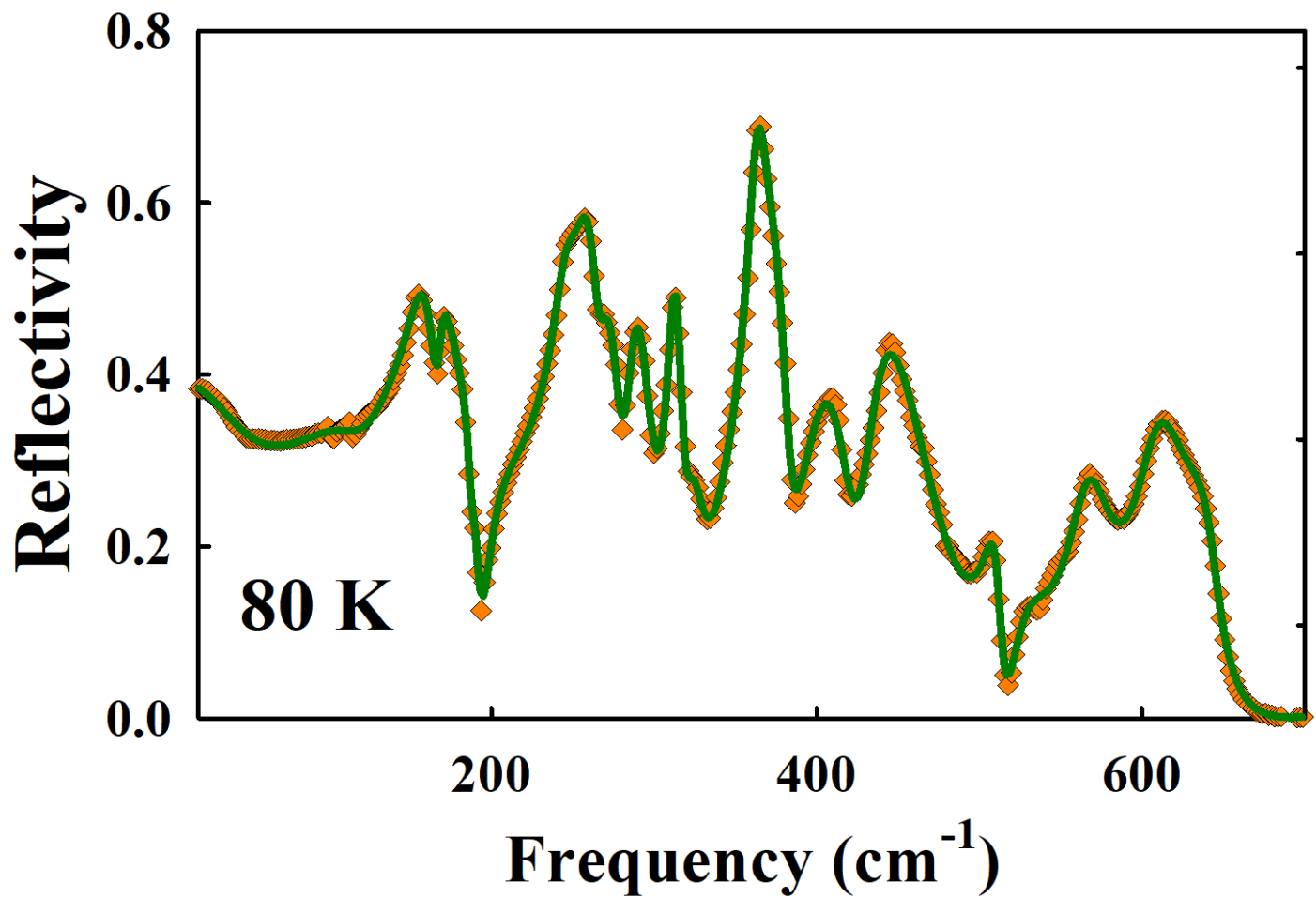

**Fig. S14.** ErFeO$_3$ near normal reflectivity at 80 K experimental: diamonds (showing 1 out of 10 points), full line: fit.

# TABLE SIII

Dielectric simulation fitting parameters for ErFeO$_3$ reflectivity at 80 K. Bottom cells show the parameters used for the lower frequency Gaussian fit for an incipient boson peak shown in Fig. S14.

| T (K) | $\varepsilon_\infty$ | $\omega_{TO}$ (cm$^{-1}$) | $\Gamma_{TO}$ (cm$^{-1}$) | $\omega_{LO}$ (cm$^{-1}$) | $\Gamma_{LO}$ (cm$^{-1}$) |
|---|---|---|---|---|---|
| 80 | 1.22 | 116.0 | 49.3 | 120.1 | 42.5 |
| | | 157.0 | 16.9 | 160.8 | 21.3 |
| | | 162.5 | 35.4 | 166.9 | 7.3 |
| | | 168.6 | 6.7 | 172.0 | 34.8 |
| | | 173.4 | 47.5 | 187.6 | 2.7 |
| | | 187.9 | 2.99 | 193.0 | 9.4 |
| | | 197.5 | 27.7 | 206.8 | 85.7 |
| | | 244.5 | 14.7 | 250.2 | 21.5 |
| | | 256.0 | 17.0 | 265.0 | 12.3 |
| | | 269.0 | 14.2 | 279.5 | 14.1 |
| | | 286.2 | 11.4 | 299.3 | 22.0 |
| | | 310.6 | 8.7 | 317.5 | 9.3 |
| | | 322.1 | 13.6 | 327.3 | 20.4 |
| | | 358.9 | 15.3 | 366.0 | 17.3 |
| | | 360.4 | 11.5 | 382.4 | 13.2 |
| | | 384.1 | 33.9 | 388.4 | 95.6 |
| | | 399.8 | 45.4 | 420.6 | 24.1 |
| | | 432.7 | 22.7 | 475.2 | 71.0 |
| | | 509.2 | 19.5 | 513.2 | 8.1 |
| | | 522.4 | 22.3 | 524.1 | 44.3 |
| | | 561.8 | 27.2 | 573.2 | 36.1 |
| | | 602.4 | 354 | 617.1 | 40.6 |
| | | 640.3 | 85.7 | 642.7 | 13.9 |
| | | 3103.6 | 4753.4 | 4457.9 | 9553.0 |
| | | A | Xc | w | |
| | | 435.4 | 7.4 | 26.1 | |

## LuFeO$_3$ -Reflectivity and Multioscillator Fitting Parameters

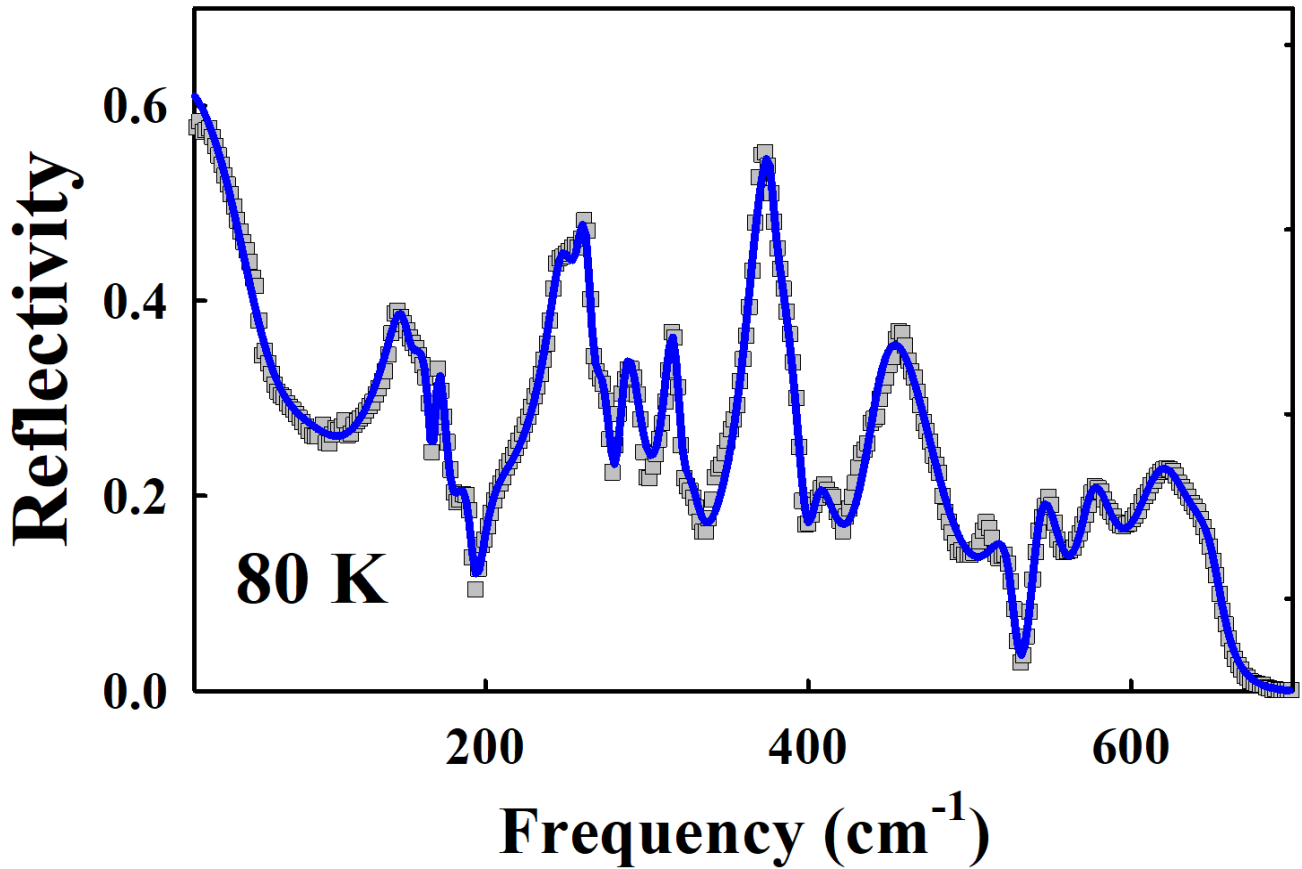

**Fig. S15.** LuFeO$_3$ near normal reflectivity at 80 K; experimental: squares (showing 1 out of 10 points), full line: fit. Note the increase in reflectivity toward lower frequencies indicates a boson peak known to be associated to structural inhomogeneity, unreleased strains, and potential freer charges localized in grain boundaries or domain walls. [10]

# TABLE SIV

Dielectric simulation fitting parameters for LuFeO$_3$ reflectivity at 80 K. Bottom cells show the parameters used for the lower frequency Gaussian fit for the boson peak shown in Fig. S15.

| T (K) | $\varepsilon_\infty$ | $\omega_{TO}$ (cm$^{-1}$) | $\Gamma_{TO}$ (cm$^{-1}$) | $\omega_{LO}$ (cm$^{-1}$) | $\Gamma_{LO}$ (cm$^{-1}$) |
|---|---|---|---|---|---|
| 80 | 1.14 | 108.7 | 90.5 | 116.2 | 64.5 |
| | | 148.5 | 23.6 | 155.1 | 5.7 |
| | | 160.5 | 6.6 | 163.8 | 21.8 |
| | | 169.4 | 6.1 | 177.3 | 13.3 |
| | | 187.6 | 21.2 | 193.4 | 10.0 |
| | | 245.9 | 15.7 | 253.7 | 18.2 |
| | | 259.6 | 12.2 | 269.4 | 11.7 |
| | | 269.8 | 16.8 | 280.2 | 133.8 |
| | | 285.2 | 11.0 | 298.2 | 29.8 |
| | | 314.2 | 11.7 | 321.0 | 10.9 |
| | | 325.1 | 14.1 | 329.4 | 21.2 |
| | | 365.8 | 18.7 | 377.7 | 8.9 |
| | | 368.2 | 83.0 | 374.2 | 61.5 |
| | | 377.2 | 10.4 | 398.4 | 13.8 |
| | | 402.7 | 14.1 | 415.4 | 43.5 |
| | | 440.7 | 29.3 | 486.3 | 68.7 |
| | | 522.7 | 31.4 | 528.5 | 11.5 |
| | | 540.2 | 13.6 | 553.7 | 33.7 |
| | | 571.7 | 22.0 | 581.5 | 35.9 |
| | | 608.8 | 42.1 | 630.5 | 41.8 |
| | | 648.9 | 43.8 | 652.3 | 18.0 |
| | | 3997.7 | 12126.6 | 5551.0 | 25947 |
| | | A | Xc | w | |
| | | 3228.1 | 10 | 24.0 | |

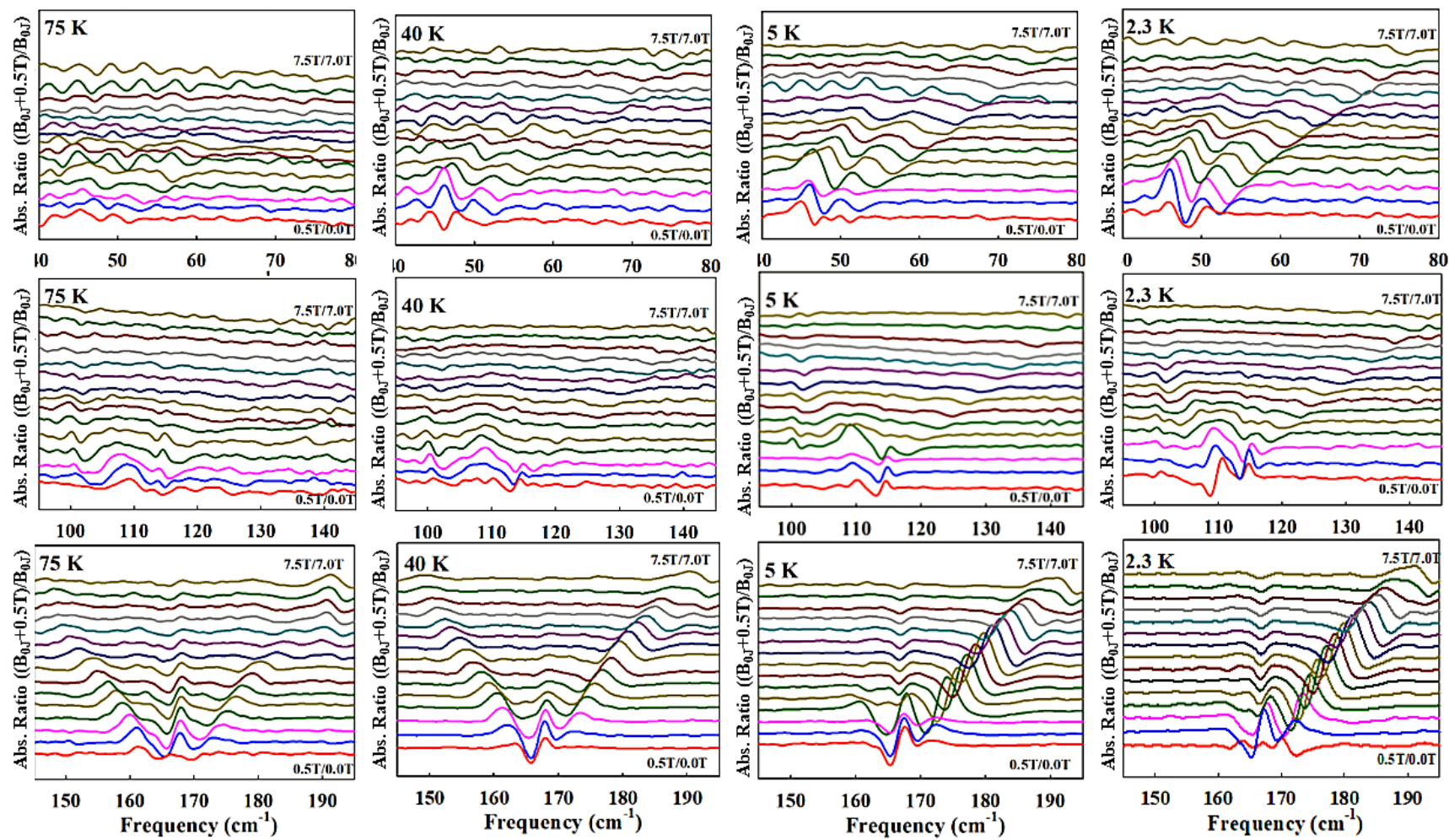

**Fig. S16.** Polyethylene pellet ErFeO$_3$ embedded sequential absorption ratio spectra, $(B_{0j}+0.5T)/B_{0j}$, $B_{0j}$ is the applied field at the j$^{th}$ incremental step, of ground crystal field multiplet Er$^{3+}$ ($^4I_{15/2}$) zero field cooled at 46 cm$^{-1}$, 114 cm$^{-1}$ and 168 cm$^{-1}$ in the Er$^{3+}$ paramagnetic and antiferromagnetic phases showing the biased distortion due to the Fe$^{3+}$ ($^6A_1$) magnetic exchange below the compensation temperature ~40 K. Spectral noise is understood in terms of the population temperature dependence according Boltzmann statistics for the respective levels.